\documentclass{article}

\usepackage[utf8]{inputenc} 
\usepackage[T1]{fontenc}    
\usepackage{hyperref}       
\usepackage{url}            
\usepackage{booktabs}       
\usepackage{amsfonts}       
\usepackage[margin=2cm]{geometry}
\usepackage{tabu}
\usepackage{tikz}
\usetikzlibrary{intersections, arrows, positioning, shapes, fit, calc, trees, backgrounds}
\usepackage{amsmath}
\usepackage{subcaption}

\title{Investigating cognitive ability using action-based models of structural brain networks}

\author{
 Viplove Arora, 
 Enrico Amico, 
 Joaqu\'in Go\~ni, and 
 Mario Ventresca \\
}

\begin{document}
\maketitle

\begin{abstract}
Recent developments in network neuroscience have highlighted the importance of developing techniques for analyzing and modeling brain networks. A particularly powerful approach for studying complex neural systems is to formulate generative models that use wiring rules to synthesize networks closely resembling the topology of a given connectome. Successful models can highlight the principles by which a network is organized (identify structural features that arise from wiring rules versus those that emerge) and potentially uncover the mechanisms by which it grows and develops. Previous research has shown that such models can validate the effectiveness of spatial embedding and other (non-spatial) wiring rules in shaping the network topology of the human connectome. In this research, we propose variants of the {\it action-based model} that combine a variety of generative factors capable of explaining the topology of the human connectome. We test the descriptive validity of our models by evaluating their ability to explain between-subject variability. Our analysis provides evidence that geometric constraints are vital for connectivity between brain regions, and an action-based model relying on both topological and geometric properties can account for between-subject variability in structural network properties. Further, we test correlations between parameters of subject-optimized models and various measures of cognitive ability and find that higher cognitive ability is associated with an individual's tendency to form long-range or non-local connections.
\end{abstract}


\section{Introduction}
\label{sec:intro}
The network of connections between neural elements of the human brain, often referred to as the human connectome \cite{Sporns2005, Sporns2010}, creates an intricate and complicated structural network \cite{y1995histology, swanson2012brain}. The human connectome is an anatomical network, where nodes consist of neural elements (neurons or brain regions), and edges correspond to physical connections (synapses or axonal projections) between different neural elements. The network map of the human connectome can be used to describe the organization of the brain's structural connections and their role in shaping functional dynamics \cite{Sporns2011, Bassett2018}. In the past decade, the areas of network neuroscience and brain connectomics have highlighted the importance of developing a wide variety of techniques for analyzing and modeling brain networks \cite{Bullmore2009, Sporns2011, Telesford2011a, Fornito2013, Sporns2014a, Sporns2018}. For instance, topological analysis based on various network measures has provided evidence for the non-random topology of the connectome, and has aided our understanding of the organization of the human brain \cite{Rubinov2010, Bassett2017}. In this research, we propose variants of the {\it action-based model} that combine a variety of generative factors capable of explaining the topology of the human connectome. Our goal is to study the role of spatial constraints and multiple wiring rules as generative rules for connectivity between brain regions and their ability to capture between-subject variability in structural network properties.

Early applications of Graph Theory and Network Science in Neuroscience mainly focused on gathering summary quantities in an attempt to find common features describing the organization of most biological neural networks \cite{Sporns2005, Bullmore2009, Sporns2011, Bullmore2011, Sporns2012a, Betzel2017, Bassett2017}, see \cite{Rubinov2010} for a review. These summary features and measurements have been used to detect functional integration (shorter path lengths and efficiency) and segregation (high transitivity and presence of clusters) in the brain. The importance of specific brain regions and pathways can be computed using centrality metrics, such as betweenness, closeness, etc \cite{Rubinov2010}. These measures have also been useful for topological analysis and characterizing structural patterns observed in the network representation of the brain \cite{Bassett2017, Sporns2018}.

\begin{figure}[!ht]
    \centering
    \begin{tikzpicture}[]
        \node[inner sep=0pt,above right, label=below:Observed Network $G^*$] (net) {\includegraphics[width=0.2\linewidth]{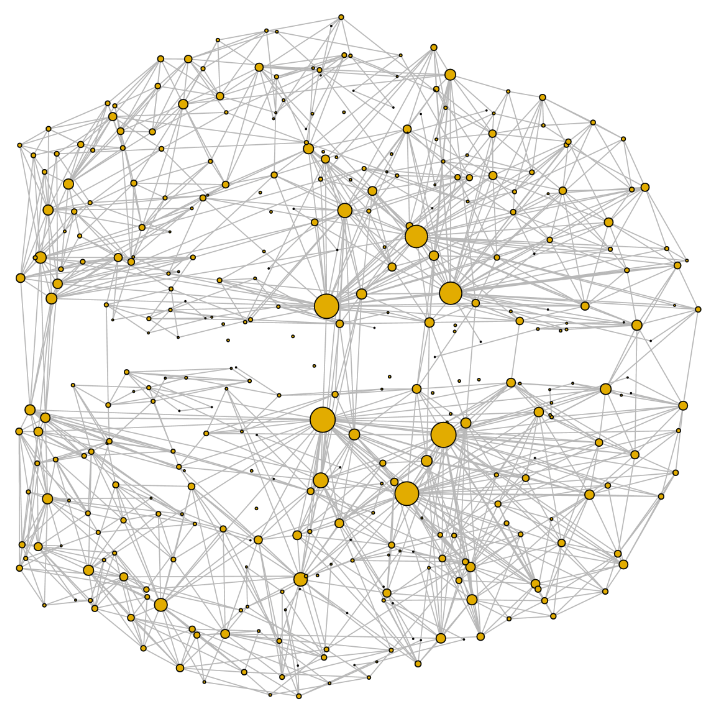}};
        \node[inner sep=0pt,right=2cm of net] (gen) {\includegraphics[width=0.35\linewidth,trim={0 450 100 0},clip]{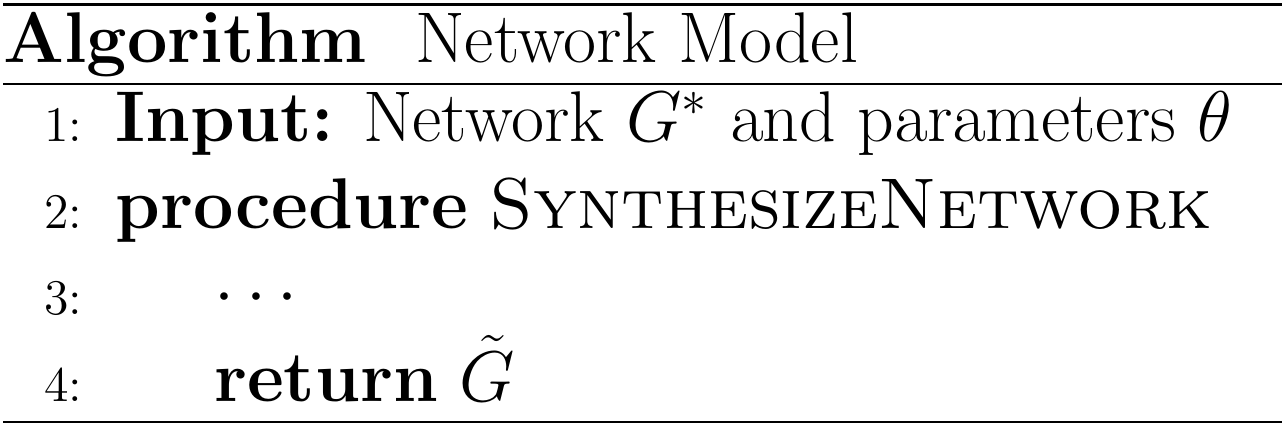}};

        \draw[->,shorten >= 5pt, thick, color=blue] (net) .. controls +(3,1.8) .. node[label=above:fit/learn $\pmb{\theta}$] {} (gen);
        \draw[->,shorten >= 5pt, thick, color=blue] (gen) .. controls +(-3,-1.8) .. node[label=below:$\pmb{\theta} \Rightarrow$ insights] {} (net);
    \end{tikzpicture}
    \caption{A generative network model uses an undirected and binarized network representation of the brain as an input to learn model parameters $\pmb{\theta}$, which can then be used to draw insights about the topology of networks similar to $G^*$.}
    \label{fig:mods_brain}
\end{figure}

An alternative approach for studying brain networks is to formulate generative models that use wiring rules to synthesize networks resembling the topology of a given connectome \cite{Simpson2012, Vertes2012, Nicosia2013a, Klimm2014, Betzel2016, Obando2017, Betzel2017} (see Figure \ref{fig:mods_brain} for a pictorial description). Those models can highlight the principles by which a network is organized (identify structural features that arise from wiring rules versus those that emerge) and potentially uncover the mechanisms by which it grows and develops \cite{Bassett2018}. For example, the spatial embedding of the brain \cite{Karbowski2001}, along with the economical wiring constraints \cite{Laughlin1998, Achard2007, Raj2011, Hahn2015, Budd2015} that arise from this embedding play a vital role in crucial network characteristics, such as efficient network communication and information processing \cite{Bassett2017}. Previous research has shown that generative models can validate the effectiveness of spatial embedding and other (non-spatial) wiring rules in shaping the network topology of the human connectome \cite{Vertes2012, Betzel2016}.

Generative models can also provide insights into potential mechanisms that give rise to functionally important network attributes \cite{Avena-Koenigsberger2014}. In addition to providing explanations for the wiring rules and processes of network formation, generative models can compress our descriptions of the network representation of the brain and highlight potential regularities in their structural organization \cite{Betzel2017}. This compact description of the wiring rules enables these models to make out-of-sample predictions about unobserved network data.

Early work on generative modeling of the human brain (using resting state functional connectivity \cite{Vertes2012} or the connectome \cite{Betzel2016}) utilized at most one or two generative factors (such as homophily, preferential attachment, and geometric distances) for synthesizing the network topology. Consequently, there has been growing interest in developing generative models that incorporate multiple rules for the probability of connections between regions of the brain \cite{Wijk2010, Klimm2014}. The action-based model for networks \cite{Arora2017a} combines several pre-defined actions/wiring rules to learn a probabilistic model for a given input network $G^*$.

The representation of complex systems using interactions and dependencies between units has been the main focus of network science \cite{Newman2003, torres2021and}. Identifying universal mechanisms for network growth that can explain the emergent patters observed in real-world networks is often cited as an important challenge in the field \cite{Stumpf2012, Holme2019}. Consequently, mechanisms like preferential attachment \cite{Barabasi1999a} and homophily \cite{McPherson2001} have been identified as driving forces for interactions in a network. These mechanism coupled with the observation that stochastic local interactions give rise to emergent global structure \cite{Ladyman2013a} has inspired researchers to combine multiple mechanisms for link formation in networks \cite{Bailey2014, Zhang2015, Arora2017a, Attar2018a, langendorf2021empirically, xiao2021deciphering}. 


The ability of a generative model to make out-of-sample predictions means that they can be utilized to synthesize networks that can capture the difference between subjects \cite{Rolnick2019}. The ability of a generative model to synthesize realistic networks and creating a compressed model of the connectome highlights the possibility of providing important insights into the factors that have shaped the emergence of specific architectural or performance characteristics of an individual \cite{Sporns2018}. This distinctive feature allows us to answer questions about how the structural organization of a brain can carry information useful for deciphering inter-individual differences in cognition \cite{Gabrieli2015, Bzdok2019}. Although little is known about the organization principles that lead to individual differences between the connectomes, it is widely believed that these differences are associated with cognitive functioning \cite{Mueller2013, Medaglia2015a, Bassett2018, Barbey2018}. Consequently, a recent topic of interest in neuroscience has been to uncover how individual differences in the network architecture are associated with differences in general intelligence and cognitive functioning \cite{Barbey2018, Tompson2018}.

Cognitive ability has usually been linked with an individual's ability to reason, learn from experience, solve problems, and think abstractly \cite{Gottfredson1997}. The $g$ factor (also known as general intelligence) proposed by Charles Spearman \cite{spearman1927abilities} is often used for measuring cognitive ability. John Carroll proposed the three-stratum theory of cognitive ability as an expansion and extension of previous theories \cite{Carroll1993}. The three-stratum theory proposes that individual differences in cognitive ability can be classified into three different strata -- narrow, broad, and general abilities \cite{carroll2005three}. With the advancement of brain imaging technologies and the availability of large datasets, computational models have been playing an important role in advancing our understanding of cognitive science \cite{Kriegeskorte2018}.

In an attempt to understand the differences in cognitive ability between individuals, researchers have tried to investigate the role of connectivity patterns between different sets of brain areas \cite{Medaglia2015a}. Using fMRI measurements, \cite{Simpson2012, Mueller2013} found that the inter-individual variability in positively associated with long-range connectivity, i.e., the differences between individuals are more pronounced in the connectivity between brain regions that are farther apart in space. Further, the individual differences in cognition can predominantly be attributed to regions with high connectivity variability \cite{Mueller2013}. Similar differences can also be observed in the anatomical or structural connectivity, which acts as a backbone for communication between different brain regions that support a wide range of cognitive functions \cite{Park2013, Misic2016, Kriegeskorte2018}. Thus, one would expect a generative model for the anatomical connectivity that can accurately deduce the individual difference in connectivity, especially in long-range connectivity, can prove useful in our quest for understanding the relation between the structural organization of the brain and the individual differences in cognitive ability. 

In this paper, we propose and assess four generative models (details in Section \ref{sec:methods_ch6}): (i) null model based only on geometric distances, (ii) action-based model proposed in \cite{Arora2016, Arora2017a}, (iii) a variant of (ii) with an additional action based on geometric distances, and (iv) action-based model with visibility, where wiring rules use both topological and geometric properties to create edges. For each of these models, we cross-validate their ability to explain between-subject variability when trained on a single group representative network $G^*$ in Section \ref{sec:mod-cv}. Following the evaluation of the four models, we use our best model, ABNG (vis), for understanding the relationship between the structural organization of human brains and the cognitive ability of subjects in Section \ref{sec:cog}. 

\section{Methods}
\label{sec:methods_ch6}
The choice of input network $G^*$ is particularly critical for generative modeling because the network should typify the complex structure of the entire set of brain networks of interest \cite{Wijk2010, Simpson2012, Klimm2014}. In this paper, we use the whole-brain structural connectivity networks of 100 unrelated subjects from the HCP dataset \cite{VanEssen2013} to create a group representative median network $G^*$ (See Section \ref{sec:brain_data} for more details). We also learn action-based models and its geometric counterpart, ABNG (vis), for each subject to study correlations between measures of cognitive ability and model parameters.

To develop an understanding of the structural organization of the human brain, we implement four generative models (briefly described in Sections \ref{sec:null}-\ref{sec:abn-vis}) that account for different mechanisms for link creation. For each of the generative models, we formulate the problem of determining parameters $\pmb{\theta}$ as a multi-objective optimization problem:

\begin{equation}
\label{eq:genform1}
\begin{aligned}
& {\text{minimize}} & & \mathbb{E}\left[Q(G|G^*, Y, \pmb{\theta})\right]\\
& \text{subject to} & & \pmb{\theta} \in D,
\end{aligned}
\end{equation}

where $G$ is a network synthesized by a generative model with parameters $\pmb{\theta}$ in the feasible domain $D$, and $Q(G|G^*, Y, \pmb{\theta})$ is a measure to quantify the dissimilarity between a synthesized network $G$ and the group representative network $G^*$ based on a user-defined set of network characteristics $Y$. We minimize the expectation of $Q$ to account for the stochasticity in the networks synthesized by a generative model. We would like to note that synthesizing graphs isomorphic to $G^*$ is the optimal but degenerate solution to the problem stated in Equation \ref{eq:genform1}. Our goal is not to exactly reproduce $G^*$ but rather learn a model that can synthesize networks statistically similar to one another and $G^*$ i.e. variation is expected/desired. Also, since generative models are inherently stochastic in nature, it is unlikely that the networks synthesized for a fixed parameters setting will be isomorphic to each other or $G^*$.

Recent observations have highlighted the need to consider multiple global characteristics when comparing networks \cite{Przulj2007, Roy2012, Yaveroglu2014a, Arora2017a}. For our experiments, we use the first three terms of the $dk$-series \cite{Orsini2015} (i.e., $Y=$ degrees $+$ correlations $+$ clustering/transitivity) as they have been shown to almost fully define local and global organization of most real-world networks. The 2-sample Kolmogorov-Smirnov (KS) $D$-statistic is used to quantify the difference in distribution of these properties between $G$ and $G^*$. As the resulting problem is multi-objective in nature, we obtain a set of Pareto efficient solutions after solving the optimization problem described in Equation \ref{eq:genform1}. For each of the models, the solution closest to the origin, i.e. the one with lowest sum of objectives based on 1-norm, was chosen as the representative parameter setting. This choice assumes that all the objectives are equally important for choosing a model, but our approach allows for the user to specify other criteria for selecting a solution of their choice from the Pareto Front.

\subsection{Null model}
\label{sec:null}
The most basic model we consider in our experiments assumes that the probability of connection $P_{ij}$ between nodes $v_i$ and $v_j$ is a function of the Euclidean distance $d_{ij}$ between them

\begin{equation}
    P_{ij} \propto \exp(- \eta d_{ij}).
    \label{eq:null}
\end{equation}

This model assumes that the topology of the connectome can be attributed to minimization of the wiring cost, and the parameter $\eta \geq 0$ can be optimized (as formulated in Equation \ref{eq:genform1}) to determine the degree of cost penalization. NSGA-II \cite{Deb2002} was used to solve the optimization problem for the null model, which resulted in $\eta \approx 0.73$ as the most representative solution. As illustrated in Figure \ref{fig:res}, networks synthesized using the estimated parameter setting for the null model were unable to match the between-subject variability in the topological properties.

\subsection{Action-based model}
\label{sec:abng_ch6}
The action-based model \cite{Arora2017a, Arora2016} uses a first principles perspective to create an algorithmic procedure that learns a compact probabilistic representation of a given input network. In the action-based model (ABNG), we assume that the generative process is composed of two main components: (i) node inherent potential to create links using different strategies capturing the collective behavior of nodes, and (ii) an algorithmic environment $F(\cdot)$ that provides opportunities for nodes to create links; thus simulating the emergence of a macroscopic structure from individual interactions. The different actions for creating a link between nodes $v_i$ and $v_j$ are enumerated using a pre-defined action set $A=\{a_1, \dots, a_K\}$, for example


\[A =\mathsf{\left\{\begin{tabular}{l}
    $a_1 =$ probabilistically select $v_j$ based on its degree,\\
    $a_2$ = probabilistically select $v_j$ based on its betweenness,\\
    $a_3 =$ select $v_j$ based on inverse log-weighted similarity to $v_i$,\\
    $a_4 =$ do not make an edge
    \end{tabular}\right\}}\]

The environment to create links using these actions is specified by a generative algorithm $F(\mathbf{M}, n)$ that can be used to synthesize networks containing $n$ nodes using an action matrix $\mathbf{M}$. 
The action matrix $\mathbf{M}=[\mathbb{P}(A = a_k)|\mathbb{P}(T=t)]$ contains the probability that action $a_k \in A$ is chosen by a node using $\mathbb{P}(A = a_k)$ conditioned on the node type $t = 1,\dots, d \ll n$ chosen using $\mathbb{P}(T = t)$ (see SI for additional details). Pareto Simulated Annealing (PSA) \cite{Optimization1998} is used to solve the optimization problem of estimating an action matrix for the group representative network $G^*$. The action set used in our experiments consists of $K=8$ actions, which are listed in Table \ref{tab:am_ch6} along with the representative action matrix $\mathbf{M}_{G^*}$. It should be noted that the estimated action matrix $\mathbf{M}_{G^*}$ contains only three actions that have a probability greater than 0.05, implying that multiple mechanisms play a dominant role in the organization of the connectome. Further, the results in Figure \ref{fig:res} shows that while ABNG can characterize the variability in degree and assortativity, but it fails to capture the variability in local clustering distributions.

\subsection{Action-based model with distance}
The importance of minimizing wiring cost necessitates an action that uses the spatial embedding of the connectome to create links between two nodes. To utilize this additional geometric information, an action is added to the model described in the Section \ref{sec:abng_ch6}, where a node $v_i$ probabilistically selects $v_j$ based on Euclidean distance. This new variant, ABNG (dist), uses the optimal cost penalization learnt in the null model ($\eta \approx 0.73$) to create an action based on geometric distance between nodes.The action matrix is optimized using PSA and the learnt model is used to synthesize networks. As seen in Figure \ref{fig:res}, ABNG (dist) outperforms ABNG in capturing the distribution of properties with the help of the additional action that is chosen with probability 0.197 in the representative action matrix shown in Table \ref{tab:am_ch6}.

\subsection{Action-based model with visibility}
\label{sec:abn-vis}
Previous research on generative models for the brain have highlighted the effectiveness of combining a distance based penalty with non-geometric rules to infer the probability of connection between different regions of the brain \cite{Vertes2012, Betzel2016}. This is a phenomena observed in many spatially embedded networks that have evolved to optimize similar functional requirements -- high efficiency of information transfer between nodes at low connection cost -- or to attain ideal balance between functional segregation and integration \cite{Bullmore2009}. This observation becomes even more relevant for the brain because of the significance of long-range connectivity variability.

We propose the intuitive concept of \emph{restricted node visibility} (or visibility in general), which states that the probability of creating a link between two nodes is determined by a combination of actions and external factors intrinsic to the nodes. For example, in networks that exist in the Euclidean space (structural brain networks, transportation networks, etc.), a visibility function can be defined using node locations. In the context of ABNG, visibility can be seen as a way of skewing an action such that a node is more likely to connect with particular sets of nodes, consequently leading to the formation of communities. The idea has some similarity to network models that infer an embedding of nodes based on topology and use it to synthesize networks, see \cite[Section~4]{Barthelemy2011} for a review of these models.

To combine the effect of multiple non-geometric actions with a distance-based penalty, we propose the action-based model with visibility, where the probability of an edge between two nodes is proportional to the output of an action ($\alpha_{ij}$) scaled by the geometric distance between the nodes: 

\begin{equation}
    P_{ij} \propto \alpha_{ij} \times \exp(- \eta d_{ij}).
    \label{eq:vis}
\end{equation}

The overall idea is that the likelihood of an edge between a pair of nodes depends on a combination of their topological properties and geometric distance. We would like to point out that by setting $\eta=0$ in Equation \ref{eq:vis}, we can recover the original action-based model described in Section \ref{sec:abng_ch6}, thus making the action-based model with visibility a generalized version of the action-based model. The process of learning such a model consists of two steps: (i) estimate action matrix $\mathbf{M}_{G^*}$ using the action-based model, and (ii) estimate the visibility parameter $\eta$ for the model learnt in step (i). In our experiments, we used the action matrix optimized for ABNG followed by optimization of the visibility parameter using NSGA-II \cite{Deb2002} to learn ABNG (vis) parameters ($\eta=0.11$) for the group representative network $G^*$. The results in Figure \ref{fig:res} shows that ABNG (vis) is the best model among the ones considered here.

Figure \ref{fig:mult_act} uses a simple example to demonstrate that a model with multiple actions can improve the quality of the networks being synthesized. ABNG with sequentially increasing number of actions (from left to right in Figures \ref{fig:mult_act} (c) and (d)) gets better at reproducing the edges present in the target network\footnote{it should be noted that reproducing the edges is not the goal of generative network modeling but we do it here to present a simple illustrative example.} shown in Figure \ref{fig:mult_act}(a). The quality of the synthesized networks is evaluated using the edge overlap, which is evaluated using the formula show in Figure \ref{fig:mult_act}(b). The same improvement is also seen when the concept of visibility is introduced in ABNG, as with the same action set the networks synthesized by ABNG (vis) in (d) outperform their counterparts with the same number of actions in (c). This further cements the need for having multiple mechanisms as well as the distance-based penalty for modeling brain networks.

\setlength{\tabcolsep}{2pt}
\tikzstyle{block1} = [rectangle, minimum height=4em]
\begin{figure}[!ht]
\begin{tikzpicture}[node distance = 0cm, auto]
    \node [block1] (tarnet) at (0,0) {\includegraphics[width=.3\textwidth,trim={350 90 530 70},clip]{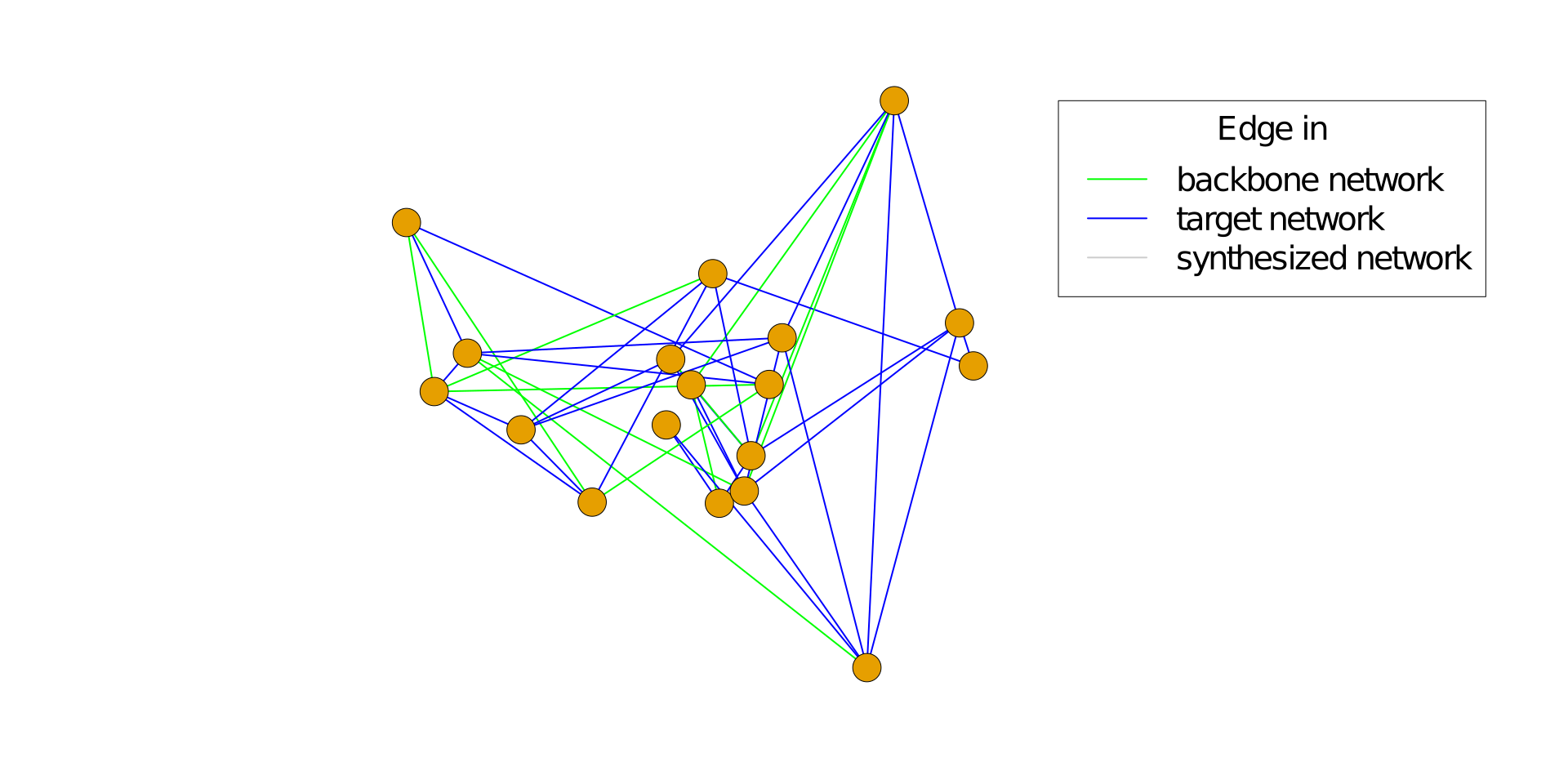}};
    \node [above left=-.5cm and -.8cm of tarnet] {{\Large \bf (a)}};
    \node [text width=25em, right=1cm of tarnet] (info) {
    \begin{center}
        \includegraphics[width=.45\textwidth,trim={950 430 60 90},clip]{tar.png}
    \end{center}
       \[\gamma = \frac{|E(G^*) \cap E(G)|}{|E(G^*)|},\]
       where $\gamma$ in the edge overlap between $G^*$ (target network) and $G$ (synthesized network), $E(\cdot)$ returns the edge set of a network, and $|\cdot|$ computes the cardinality of a set.};
    \node [above left=-.2cm and -.8cm of info] {{\Large \bf (b)}};

    \node [block1, below=.5cm of tarnet] (a1) {\includegraphics[width=.3\textwidth,trim={100 100 100 80},clip]{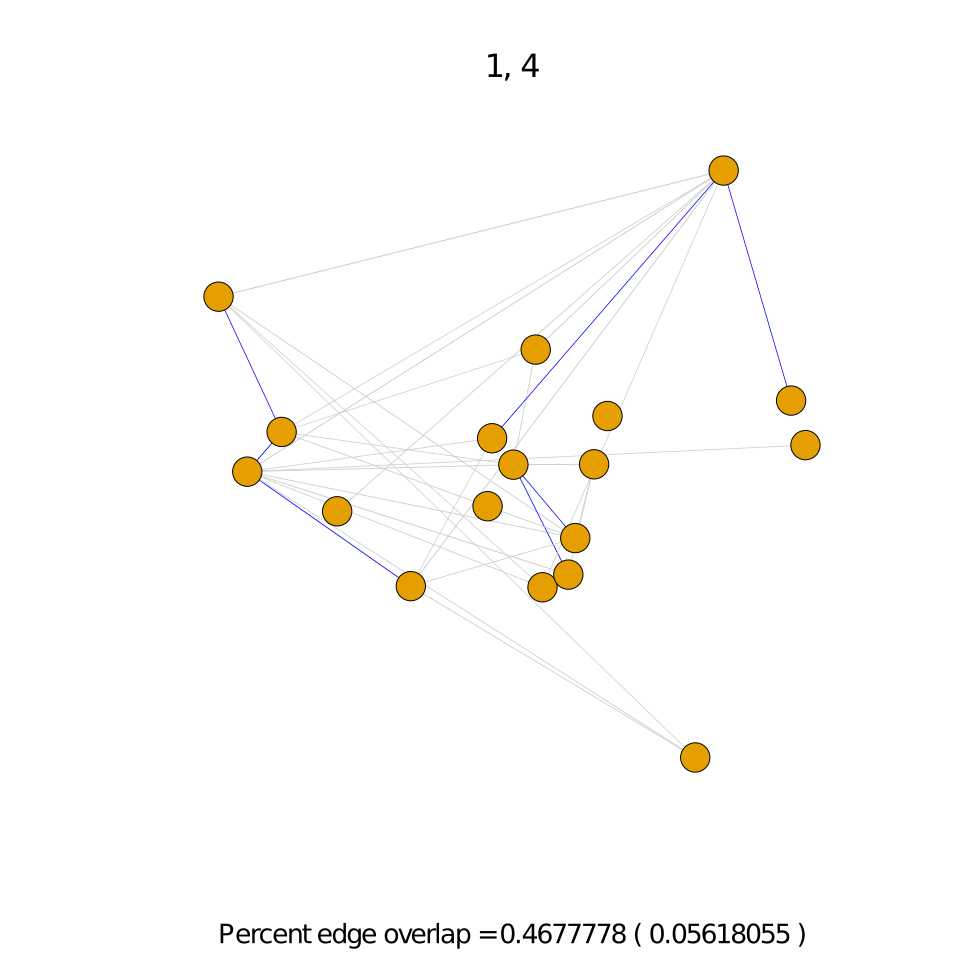}};
    \node [above=-.6cm of a1] {\begin{tabular}{ccc}
        $A=$ & $\{a_1,$  & $a_4\}$ \\ 
        $\mathbf{M=}$ & [1 & 0]
    \end{tabular}};
    \node [below left=-1.5cm and -2cm of a1] {$\bar{\gamma} = 0.47$};
    \node [block1, right= of a1] (a4) {\includegraphics[width=.3\textwidth,trim={100 100 100 80},clip]{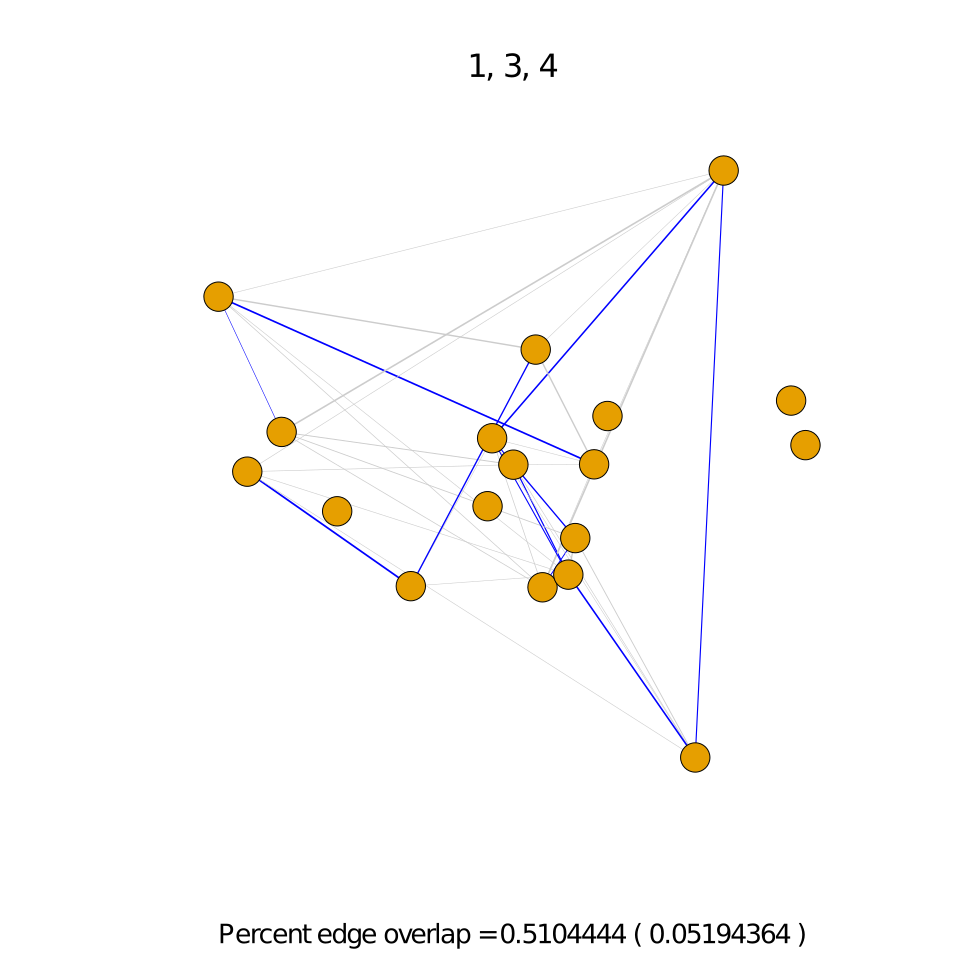}};
    \node [above=-.6cm of a4] {\begin{tabular}{cccc}
        $A=$ & $\{a_1,$ & $a_3,$ & $a_4\}$ \\ 
        $\mathbf{M}=$ & [0.1 & 0.33 & 0.57]
    \end{tabular}};
    \node [below left=-1.5cm and -2cm of a4] {$\bar{\gamma} = 0.51$};
    \node [block1, right= of a4] (a5) {\includegraphics[width=.3\textwidth,trim={100 100 100 80},clip]{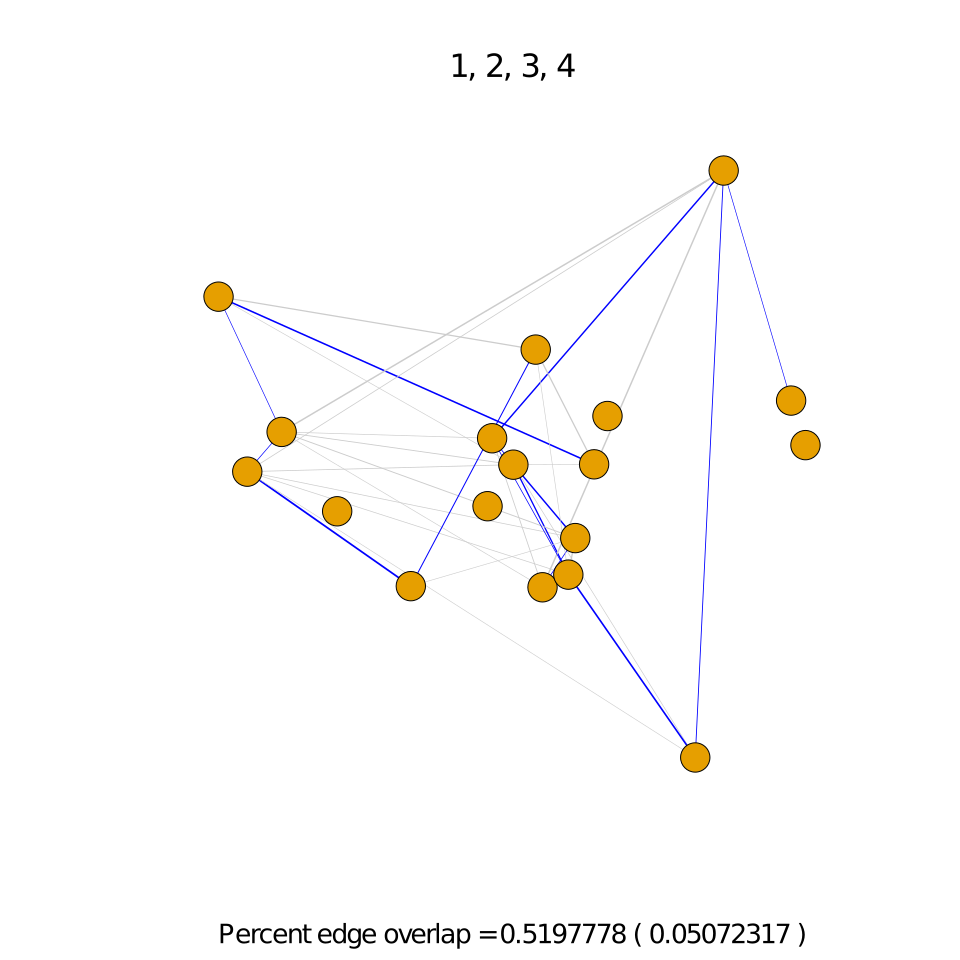}};
    \node [above=-.6cm of a5] {\begin{tabular}{ccccc}
        $A=$ & $\{a_1,$ & $a_2,$ & $a_3,$ & $a_4\}$ \\ 
        $\mathbf{M}=$ & [0.09 & 0.29 & 0.51 & 0.11]
    \end{tabular}};
    \node [below left=-1.5cm and -2cm of a5] {$\bar{\gamma} = 0.52$};
    \node [above left=-.5cm and -.8cm of a1] {{\Large \bf (c)}};

    \node [block1, below= of a1] (a1_vis) {\includegraphics[width=.3\textwidth,trim={100 100 100 80},clip]{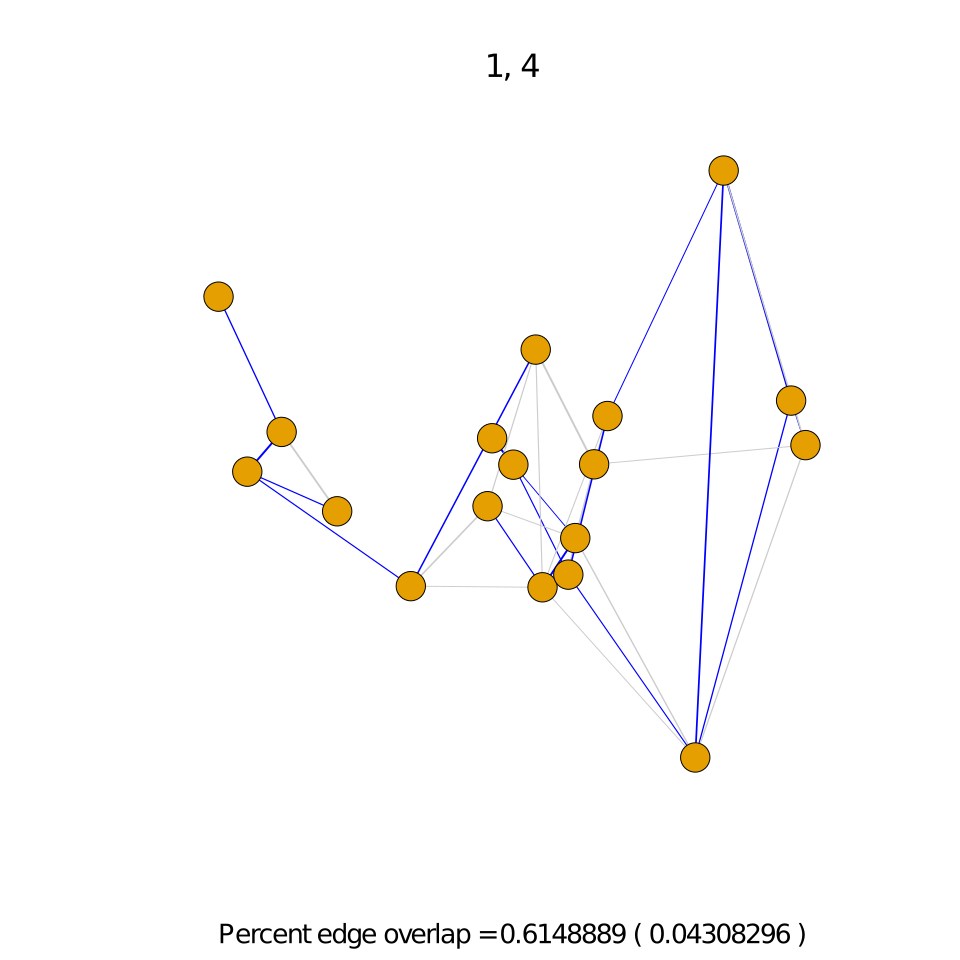}};
    \node [below left=-1.5cm and -2cm of a1_vis] {$\bar{\gamma} = 0.61$};
    \node [block1, right= of a1_vis] (a4_vis) {\includegraphics[width=.3\textwidth,trim={100 100 100 80},clip]{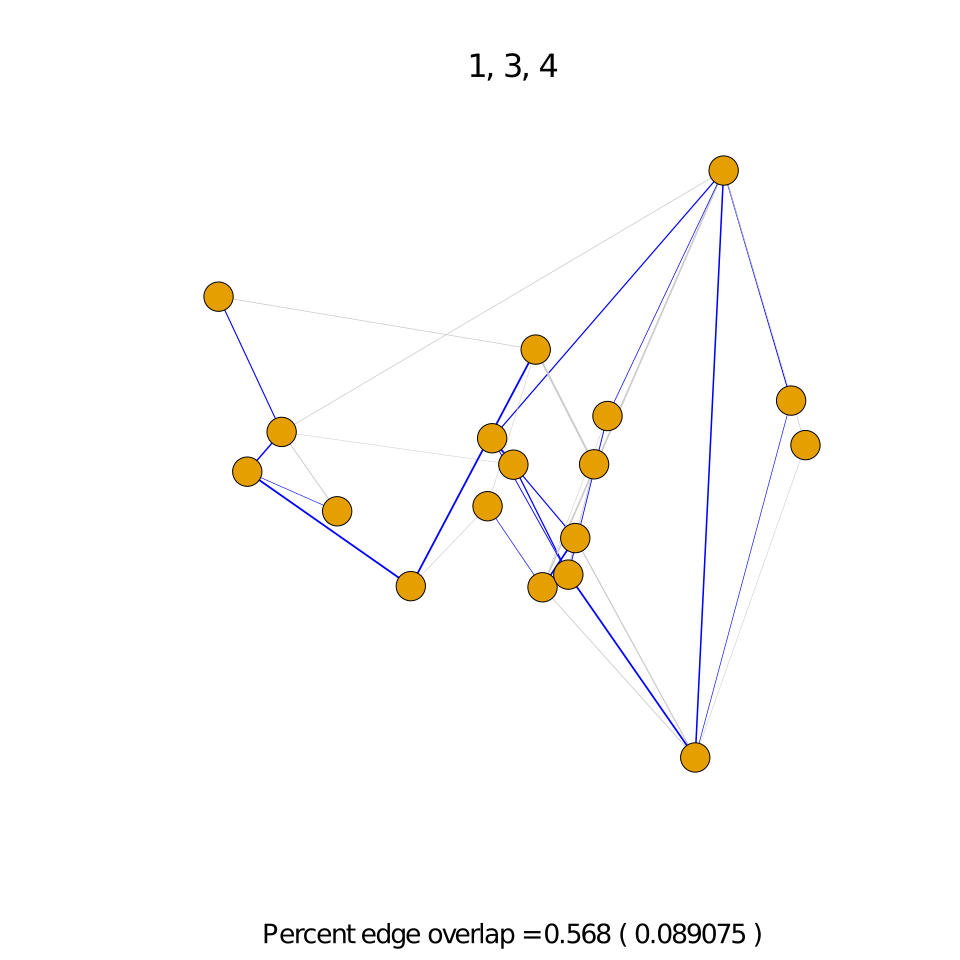}};
    \node [below left=-1.5cm and -2cm of a4_vis] {$\bar{\gamma} = 0.57$};
    \node [block1, right= of a4_vis] (a5_vis) {\includegraphics[width=.3\textwidth,trim={100 100 100 80},clip]{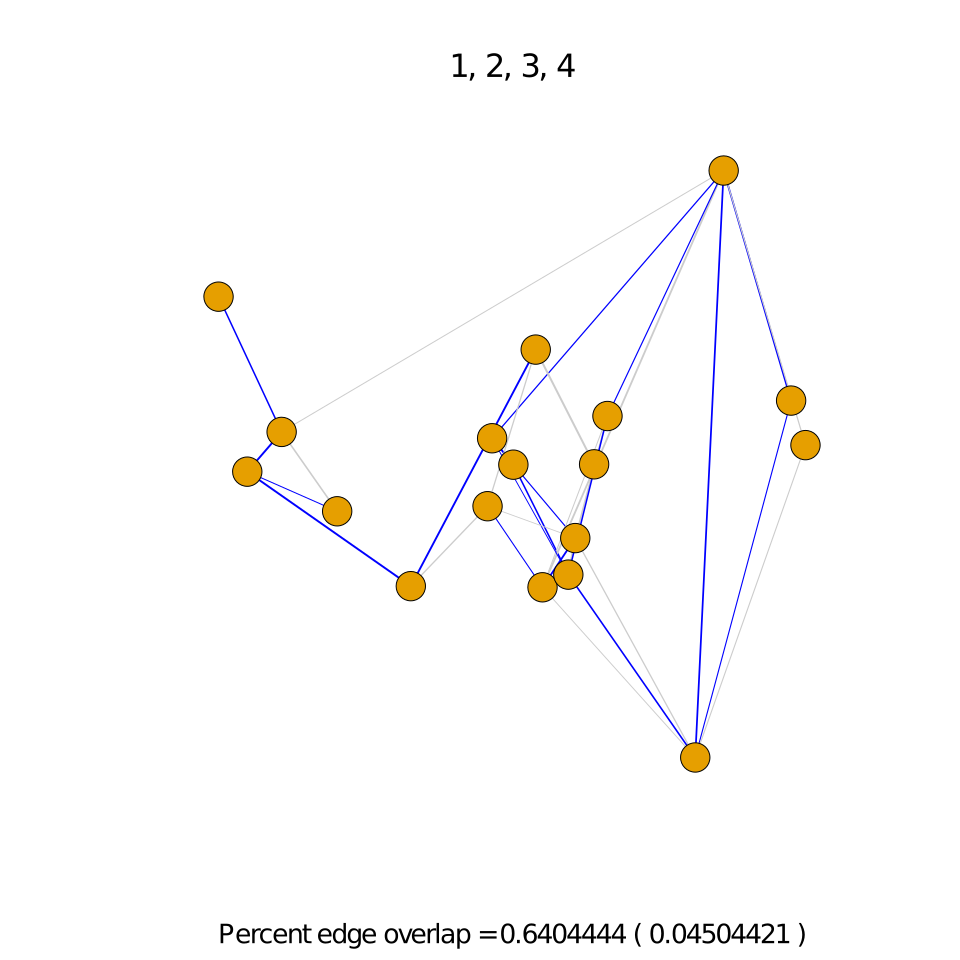}};
    \node [below left=-1.5cm and -2cm of a5_vis] {$\bar{\gamma} = 0.64$};
    \node [above left=-.5cm and -.8cm of a1_vis] {{\Large \bf (d)}};
\end{tikzpicture}
\caption{Highlighting the need for multiple mechanisms in network synthesis using a toy example: {\bf (a)} The target network $G^*$ constructed using structural brain data. Each node in this network corresponds to one of the 18 RSN (resting state network) regions. The network has 45 edges. The network constructed using the green edges corresponds to the backbone network ($G^0$) used by the action-based model. {\bf (b)} Legend describing the color coding of different edges. To test the effectiveness of using multiple mechanisms/actions in the generative model, we evaluate the edge overlap $\gamma$ between the target and synthesized networks. We use ABNG and ABNG (vis) with varying action sets $A$ and the corresponding action matrices $\mathbf{M}$ to synthesize 100 networks for each scenario. In {\bf (c)} (ABNG) and {\bf (d)} (ABNG (vis)), each network shows the 45 most likely edges generated in the 100 synthesized networks, with the edges that are also present in the target network shown in blue. For clarity, we have omitted the green edges in the backbone as they are present in every network. Mean edge overlap $\bar{\gamma}$ is also reported in every figure.}
\label{fig:mult_act}
\end{figure}


\section{Experiments and results}
To assess the validity and effectiveness of the generative models proposed in Section \ref{sec:methods_ch6}, we performed the experiments outlined in Figure \ref{fig:setup}. The first step is to create a group representative median network $G^*$ using the measurements of structural organization of the brains of the 100 unrelated subjects in the HCP dataset \cite{VanEssen2013} (see Section \ref{sec:brain_data} for further details). This is a crucial step as the models discussed in Section \ref{sec:methods_ch6}, similar to most generative network models in the literature, are designed to learn parameters using a single input network. Thus, creating an input network that can capture the structural regularities of a cohort of subjects can facilitate the learning of better models. The importance of choosing a representative network for the parameterization of generative models for the brain has also been highlighted in previous research \cite{Wijk2010, Simpson2012, Klimm2014}.

\tikzstyle{block} = [rectangle, draw, text width=5em, text centered, rounded corners, minimum height=4em, font=\footnotesize]
\tikzstyle{block1} = [rectangle, draw, text width=6em, text centered, rounded corners, minimum height=4em, font=\footnotesize]
\tikzstyle{block2} = [rectangle, draw, text width=3em, text centered, rounded corners, minimum height=4em, font=\footnotesize]
\tikzstyle{line} = [draw, -latex']
\tikzstyle{cloud} = [draw, ellipse, node distance=.7cm, font=\scriptsize, text width=50pt, text centered]
\begin{figure}[!ht]
\centering
    \begin{tikzpicture}[node distance = 0.7cm, auto]
        \node [block1] (data) {Subject 1\\ Subject 2\\ $\vdots$ \\ Subject 100};
        \node [above=.1 of data, font=\footnotesize] {Data};
        \node [block1, right=1.8 of data] (net) {\includegraphics[width=\linewidth]{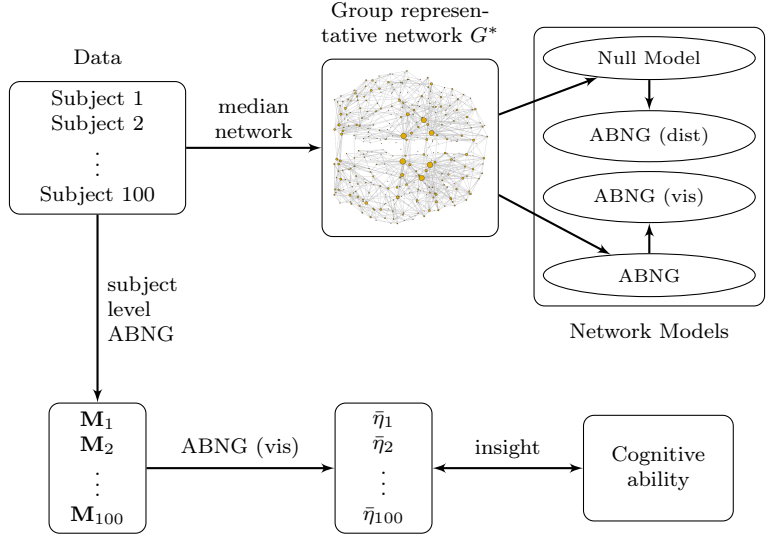}};
        \node [above=.1 of net, text width=8em, text centered, font=\footnotesize] {Group representative network $G^*$};
        \node [cloud, above right=-.2 and 1 of net] (null) {Null Model};
        \node [cloud, below right=.3 and 1 of net] (abng) {ABNG};
        \node [cloud, below=0.4 of null] (abn_dis) {ABNG (dist)};
        \node [cloud, above=0.4 of abng] (abn_vis) {ABNG (vis)};
        \node [block2, below=2.5 of data] (AMs) {$\mathbf{M}_1$\\ $\mathbf{M}_2$\\ $\vdots$\\ $\mathbf{M}_{100}$};
        \node [block2, right=2.5 of AMs] (eta) {$\bar{\eta}_1$\\ $\bar{\eta}_2$\\ $\vdots$\\ $\bar{\eta}_{100}$};
        \node [block, right=2 of eta] (GI) {Cognitive ability};
        \node [draw, rounded corners, fit= (null) (abng)] (models) {};
        \node [below=.1 of models, font=\footnotesize] {Network Models};
        \path [->, thick,line] (data) -- node[text width=5em, text centered, font=\footnotesize] {median \\ network} (net);
        \path [->, thick,line] (net) -- (null);
        \path [->, thick,line] (net) -- (abng);
        \path [->, thick,line] (null) -- (abn_dis);
        \path [->, thick,line] (abng) -- (abn_vis);
        \path [->, thick,line] (data) -- node[text width=5em, font=\footnotesize] {subject\\level\\ABNG} (AMs);
        \path [->, thick,line] (AMs) -- node[text width=6em, text centered, font=\footnotesize] {ABNG (vis)} (eta);
        \draw [>=latex' ,<->, thick] (eta) -- node[text width=5em, text centered, font=\footnotesize] {insight} (GI);
    \end{tikzpicture}
    \caption[Experimental setup for evaluating the generative models]{The structural brain networks of 100 unrelated subjects from the HCP dataset are used to create a group representative median network $G^*$. This network is used to parameterize the four models described in Section \ref{sec:methods_ch6}. Each model is cross-validated by evaluating their ability to explain between-subject variability when parameterized using $G^*$ (see Figure \ref{fig:res}). We also learn action-based models and its geometric counterpart ABNG (vis) for each subject to study correlations between measures of cognitive ability and mean model parameters $\bar{\eta}$ (see results in Section \ref{sec:cog}).}
    \label{fig:setup}
\end{figure}

Once the group representative median network is constructed, it can be used as the input to learn parameters for each of the models, as previously described in Figure \ref{fig:mods_brain} and Equation \ref{eq:genform1}. The parameterized models are then used to synthesize networks and their ability to replicate the structural features observed in the networks in the HCP dataset is evaluated in Section \ref{sec:mod-cv}. While the group representative network $G^*$ can capture the structural regularities of the cohort of subjects, it is expected that there will be subtle distinct features that are important for interpreting the difference between individuals \cite{Gordon2015}. The next step is to parameterize the models separately for each subject, and test if the fitted parameters can provide insights that can discern these individual differences. Structural brain networks are quantitative measurements of white matter micro-structure, whose integrity is crucial for healthy cognitive function \cite{Roberts2013}. Consequently, we decided to use our best model, ABNG (vis), for understanding the relationship between the structural organization of human brains and the cognitive ability of subjects in Section \ref{sec:cog}.

\subsection{Model evaluation}
\label{sec:mod-cv}
Models fitted using Equation \ref{eq:genform1} should be able to synthesize networks that replicate some properties of the group representative network $G^*$. To evaluate descriptive validity, we can use the best-fitting parameters from a model to synthesize networks that provides good estimates for the topological properties of a second network that was not involved in the model-fitting process. Such a procedure can help us ensure that the generative model is identifying wiring rules and not overfitting the observed data \cite{Betzel2017}. Some recent research has highlighted the importance of examining the variability in network populations synthesized by generative models \cite{Arora2019, Arora2020}. Following these observations, we test the ability of the network models to reproduce the topological variability across subjects in Figure \ref{fig:res}. Our results show that the different variants of the action-based model significantly outperform the null model (and other network models, see results in Section \ref{sec:res-mods}), thus showing that the proposed models can highlight potential regularities in the structural organization of the brain by using heterogeneous wiring rules. Figure \ref{fig:res} comprises of three different plots described below:

\begin{figure}[!ht]
\includegraphics[width=\linewidth]{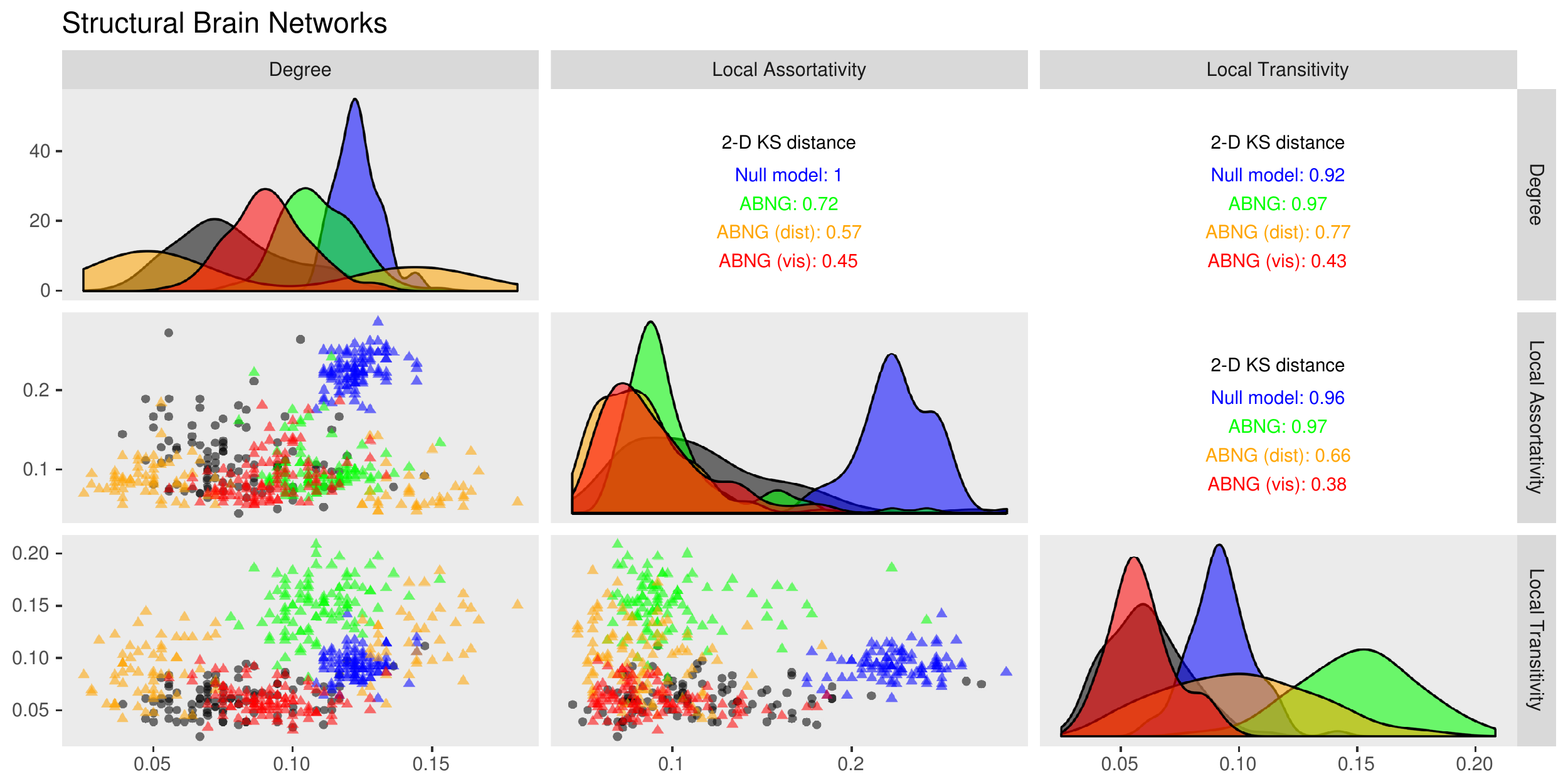}
\caption[Cross-validation of network models]{Empirical evaluation of the ability of the aforementioned network models to capture the between-subject variability using the group representative network $G^*$ as the input.}
\label{fig:res}
\end{figure}

\begin{enumerate}
\item Scatter plots below the diagonal show each synthesized/real network as a point in a network dissimilarity space, where the coordinates are computed using the KS distance of the associated properties when the network is compared to the observed network $G^*$. Network models (colored triangles) showing higher overlap with the real brain networks (black dots) are better.
\item In the blocks above the diagonal, we quantify the extent to which a given generative model is able to reproduce the between-subject variability of topological network properties of the 100 subjects using the 2-D KS distance \cite{peacock1983} (lower the better).
\item Plots along the diagonal show the density distributions of the KS distance of the associated properties when the network is compared to the observed network $G^*$. Similar density distribution to the real brain networks (black curves) implies good match in the properties.
\end{enumerate}

\begin{table}[!ht]
\centering
\caption[Learnt parameters for different models]{The table shows optimized action matrices for the group representative structural brain network $G^*$. The following actions were used: Preferential attachment on - average neighbor degree (PAND), degree (PAD), PageRank (PAPR) and betweenness (PAB); Triadic closure (TC); Inverse log-weighted (SLW) and Jaccard similarities (SJ); No action (NA); and Euclidean distance (ED). $\bar{P}$ corresponds to $\mathbb{P}(T = t)$, while $\eta$ is the optimal distance penalty parameter for each of the models. The parameters are color coded to match with Figure \ref{fig:res}: null model is blue, ABNG is green, ABNG (dist) is orange, and ABNG (vis) is red.}
\begin{tabu}{rrrrrrrrr|r|c}
\multicolumn{3}{c}{} & \multicolumn{3}{c}{Triadic closure} & \multicolumn{3}{c}{No action} & \multicolumn{2}{c}{Distance Penalty} \\
\multicolumn{4}{c}{} & \multicolumn{1}{c}{\downbracefill} & \multicolumn{2}{c}{} & \multicolumn{1}{c}{\downbracefill} & \multicolumn{2}{c}{} & \multicolumn{1}{c}{\downbracefill} \\
PAND & PAD & PAPR & PAB & TC & SLW & SJ & NA & ED & $\bar{P}$ & $\eta$\\
\hline
\rowfont{\color{green}}
0 & 0.004 &  0.013 & 0.091 & 0 & 0.492 & 0.384 & 0.016 & - & 1 & \textcolor{red}{0.111} \\
\hline
\rowfont{\color{orange}}
0 & 0 & 0 & 0.030 & 0.731 & 0 & 0 & 0.042 & 0.197 & 1 & \textcolor{blue}{0.731} \\
\multicolumn{4}{c}{\upbracefill} & \multicolumn{1}{c}{} & \multicolumn{2}{c}{\upbracefill} & \multicolumn{1}{c}{} & \multicolumn{1}{c}{\upbracefill}\\
\multicolumn{4}{c}{Preferential attachment} & \multicolumn{1}{c}{} & \multicolumn{2}{c}{Similarity} & \multicolumn{3}{c}{Euclidean Distance}\\
\end{tabu}
\label{tab:am_ch6}
\end{table}

The results in Figure \ref{fig:res} clearly show that the action-based model with visibility turns out be the best model, which is in agreement with past observations stating that the organization of the human brain arises from a combination of wiring rules based on wiring cost reduction and topological attachment mechanisms \cite{Vertes2012, Betzel2016}. In addition to learning accurate models for data, the parameters of our action-based approach can be used to draw conclusions about potential mechanisms for network formation. The action matrices shown in Table \ref{tab:am_ch6} suggest that multiple mechanisms might be at play in the organization of the human brain. The parameters for the action-based model show that homophilic attachment (action based on similarity of neighborhoods) mechanisms are the most important, but preferential attachment on betweenness is also crucial. Interestingly, the fitted distance penalty parameter for ABNG (vis) is smaller than the one obtained for the null model leading to a model that can better explain the individual variability.

\subsection{Cognitive ability from structural connectivity}
\label{sec:cog}
Our analysis so far has been restricted to evaluate the ability of the generative models to reproduce the between-subject variability of various topological properties, while using only a single group representative network as the input. The assumption that the connectomes of different subjects are topologically similar is pivotal to such an analysis \cite{Hinne2013}. While the group representative network $G^*$ can capture the structural regularities of the cohort of subjects, it is expected that the subtle differences in the connectivity patterns of different subjects are important for interpreting the difference between individuals \cite{Simpson2012, Klimm2014}. Because the action-based model combines a variety of generative factors (such as, preferential, homophilic, and distance-based mechanisms) capable of explaining the topology of the human connectome and also accurately models the between-subject variability, it can highlight potential regularities in the structural organization of an individual's brain. Thus, we parameterize ABNG (vis) for the brain networks of each individual subject and use the model parameters to discern the differences in the structural organization of brains of different subjects and its relation to cognitive ability.

\begin{figure}[!ht]
\centering
\begin{subfigure}[b]{0.45\textwidth}\centering
\includegraphics[width=.9\linewidth]{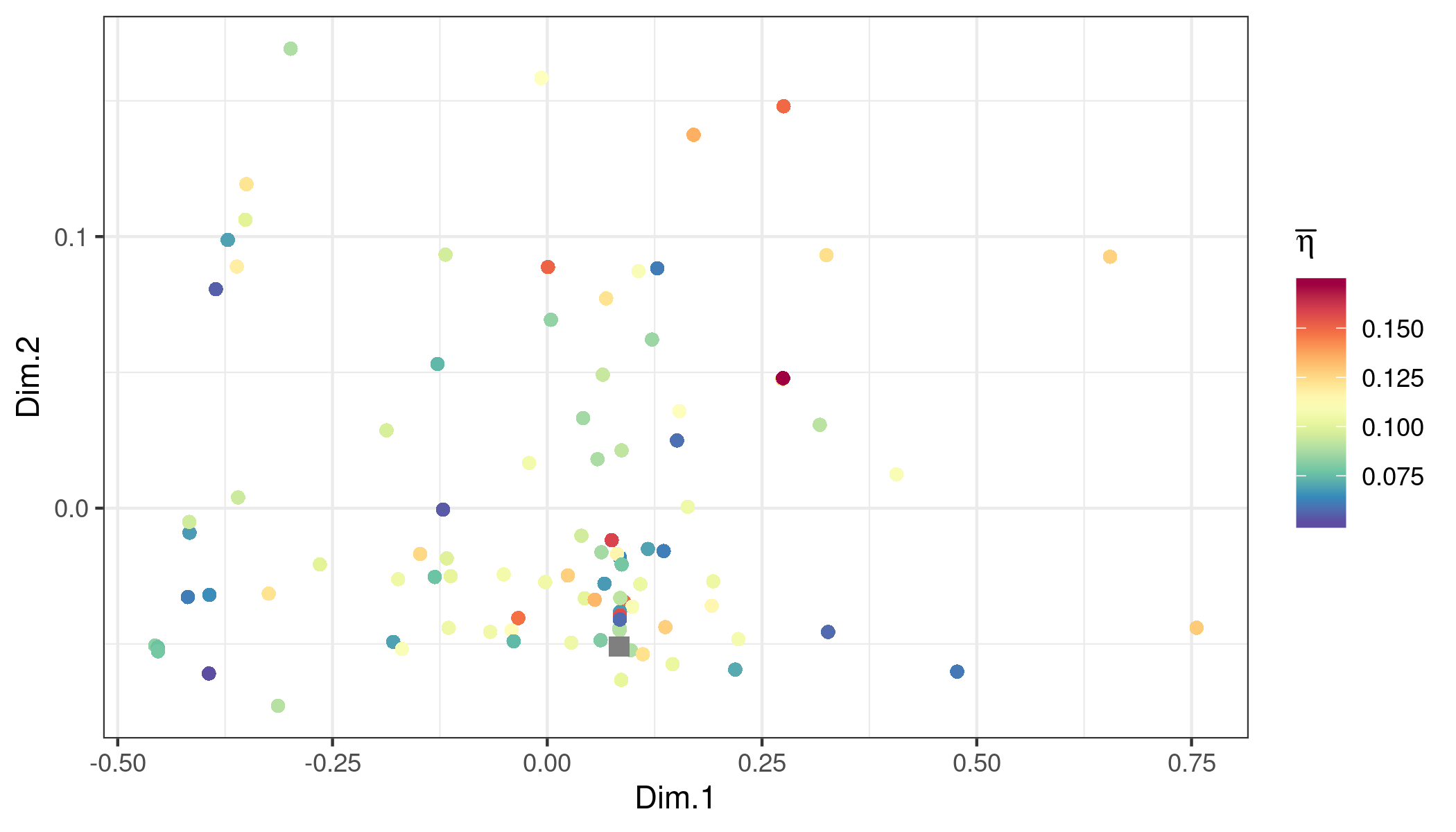}
\caption{Multi-dimensional scaling of a single representative action matrix for the 100 subjects, color shows mean visibility parameter $\bar{\eta}$.}
\label{fig:mds}
\end{subfigure}
\hspace{.2cm}
\begin{subfigure}[b]{0.5\textwidth}\centering
\includegraphics[width=.45\linewidth]{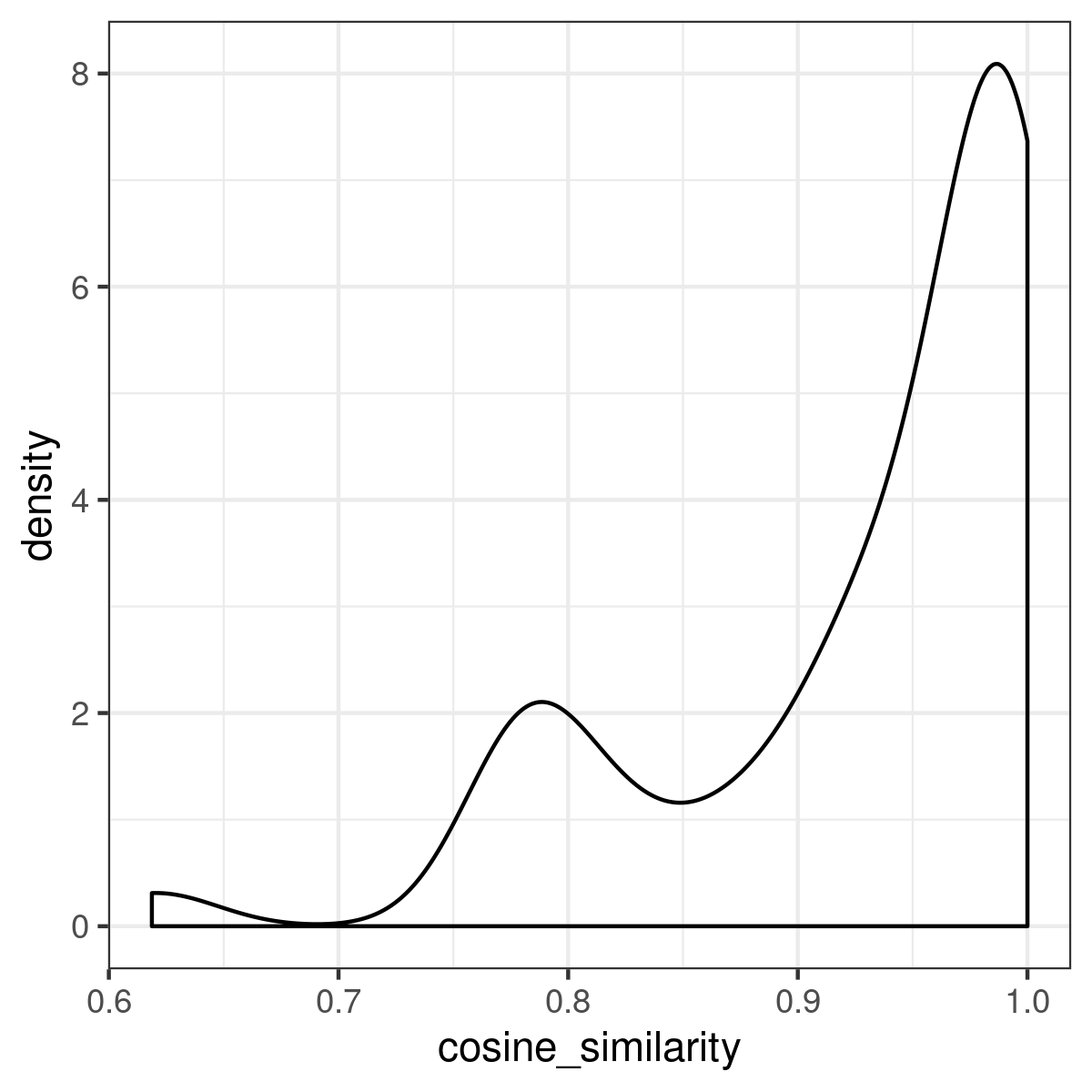} \includegraphics[width=.45\linewidth]{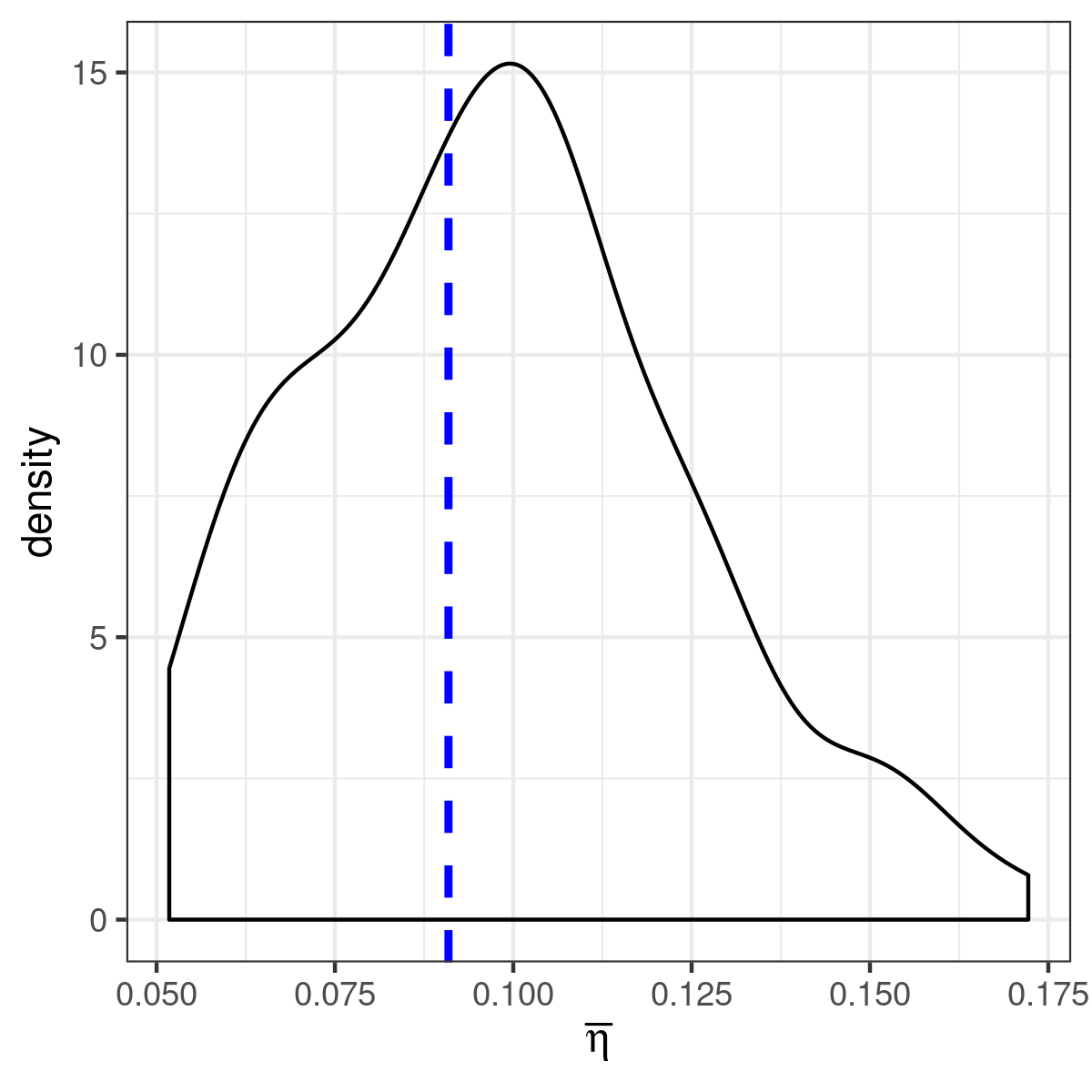}
\caption{(left) Cosine similarity between $\mathbf{M}_{G^*}$ and action matrices $\mathbf{M}_1, \dots, \mathbf{M}_{100}$ for each subject. (right) Distribution of $\bar{\eta}$ across subjects. $\bar{\eta}$ for $G^*$ is shown using the vertical blue line.}\label{fig:density}
\end{subfigure}
\caption{Visualizing the distribution of model parameters for individual subjects.}
\label{fig:AM-eta}
\end{figure}

Parameterizing ABNG (vis) for each subject involves first obtaining action matrices $\mathbf{M}_1, \dots, \mathbf{M}_{100}$ followed by estimation of the respective mean visibility parameters $\bar{\eta}_1, \dots, \bar{\eta}_{100}$ (see Figure \ref{fig:setup} for a pictorial description of our procedure). In the optimization of the action matrix for each subject, we use the action matrix $\mathbf{M}_{G^*}$ obtained for the group representative network as the starting solution (shown in green in Table \ref{tab:am_ch6} and as a grey square in the 2D space of Figure \ref{fig:mds}), and perform a local search for each subject. After obtaining the most representative action matrices $\mathbf{M}_1, \dots, \mathbf{M}_{100}$ for each subject, we perform multi-dimensional scaling of the action matrices to highlight the variation in the subject-level models. The results are shown in Figure \ref{fig:AM-eta}, where we observed that while the optimized action matrices for most subjects are similar to $\mathbf{M}_{G^*}$, there is some variability between subjects in the action matrices $\mathbf{M}_1, \dots, \mathbf{M}_{100}$ as well as the mean visibility parameters $\bar{\eta}_1, \dots, \bar{\eta}_{100}$.

A widely used measure of cognitive ability is general intelligence (first defined by Spearman \cite{spearman1927abilities} as the {\it g factor}), which is typically associated with the ability of an individual to perform a wide variety of cognitively challenging tasks well. An empirical approximation of the general intelligence of a subject can be obtained using the first principal component (PCA) of multiple measures of cognition \cite{Schultz2016}. Using the HCP dataset \cite{VanEssen2013}, we compute general intelligence using the following six measures of cognitive ability: (i) fluid intelligence (PMAT24\_A\_CR), (ii) episodic memory (PicSeq\_Unadj), (iii) cognitive flexibility (CardSort\_Unadj), (iv) language and vocabulary comprehension (PicVocab\_Unadj), (v) verbal episodic memory (IWRD\_TOT), and (vi) working memory (ListSort\_Unadj).

As discussed in the introduction, the patterns in the structural connectivity are somehow related to an individuals' general intelligence. In fact, there has been research supporting that the efficiency of network topology is positively associated with cognitive ability \cite{VandenHeuvel2009, Li2009a}. Building on the observations, we use the mean visibility parameter in ABNG (vis) as a proxy measure for the extent of functional integration and segregation in the structure of an individuals' brain, which plays a pivotal role in the ability of an individual to perform a variety of functional tasks \cite{Park2013}.

\begin{figure}[!ht]
    \centering
    \includegraphics[width=.75\linewidth]{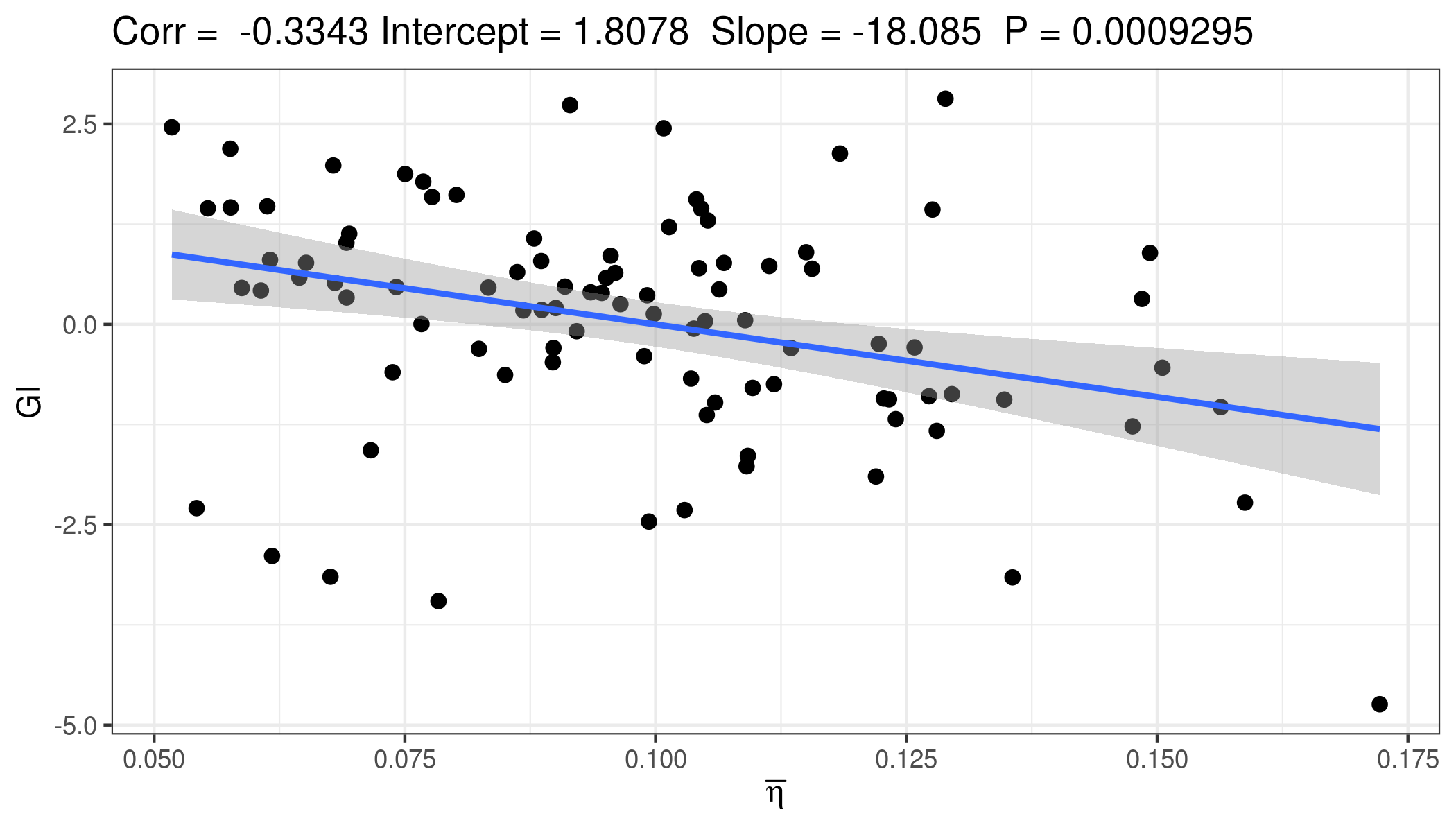}
    \caption{Testing the correlations between mean visibility parameters $\bar{\eta}$ of the individual-based models with the empirically estimated general intelligence of a subject.}
    \label{fig:GI}
\end{figure}

Figures \ref{fig:GI} and \ref{fig:cog}--\ref{fig:GI-gen} plot the mean visibility parameter $\bar{\eta}$ (averaged across the Pareto front) for an individual against different measures of cognitive ability for individual subjects. Figures \ref{fig:GI} and \ref{fig:GI-mods} plot general intelligence as the measure of cognitive ability with the mean visibility parameter $\bar{\eta}$ for three different models optimized for individual subjects. We see that ABNG (vis) with a separate AM for each subject outperforms the other models (null model for each subject and ABNG (vis) using a common AM $M_{G^*}$ for each subject), thus further highlighting the need for individually parameterized models. A common observation across all our evaluations is that individuals with lower value of $\bar{\eta}$, i.e. individuals showing a tendency to form long-range or non-local connections due to a lower distance penalty, obtain higher scores in the different tests for evaluating cognitive ability. This correlation agrees with the intuitive idea that high clustering (due to high distance penalty) favours locally specialized processing whereas short path length (due to low distance penalty) favours globally distributed processing \cite{Bullmore2009}.

\section{Conclusions}
In this paper, we explored the ability of the action-based model (and its variants) to capture the between-subject variability in topological properties of structural brain networks while using a single group representative network as the input. Though the action-based model performed better than other generative models proposed in the literature, it failed to capture the local transitivity observed in the connectome. To tackle this issue, we used the spatial embedding of the brain and introduced geometric distances between various nodes as an additional factor responsible for the topology of the connectome. This enabled us to combine multiple topological properties (in the form of different actions) and their interaction with geometric distances to create better models for the human brain, something that prior models were unable to accomplish \cite{Vertes2012, Betzel2016}. Our results show that actions-based models with geometric constraints using wiring rules based on homophilic attachment and preferential attachment on betweenness can synthesize networks resembling human connectomes.

While other generative network models such as exponential random graphs \cite{Simpson2011, Simpson2012} and the weighted stochastic block model \cite{betzel2018diversity} have also been used for generative modeling of the connectome, it remains difficult to use these models for recovering plausible mechanisms and rules that lead to the formation of an observed network. The action-based model with visibility outputs an action-matrix that shows the relative importance of various actions/mechanisms for a particular input network, and the visibility parameter highlights the role wiring cost plays in the organization of the connectome, thus providing a compact representation of the connectome.

The ability of our proposed models to synthesize networks that account for the topological properties and between-subject variability in these properties raises the possibility that the models can provide insights into the individual differences between subjects \cite{Sporns2018}. To test this hypothesis, we use our best model, ABNG (vis), to study differences in estimated parameters for different individuals and discover that the value of distance penalization is significantly correlated with cognitive ability in the form of general intelligence. We find that the differences in structural connectivity have some association with the the cognitive ability, specifically with the extent of functional integration and segregation. 

\section{Supplementary Information}

\tikzstyle{block} = [rectangle, draw, text centered, rounded corners, minimum height=4em]
\tikzstyle{block1} = [rectangle, minimum height=4em]
\tikzstyle{line} = [draw, -latex']
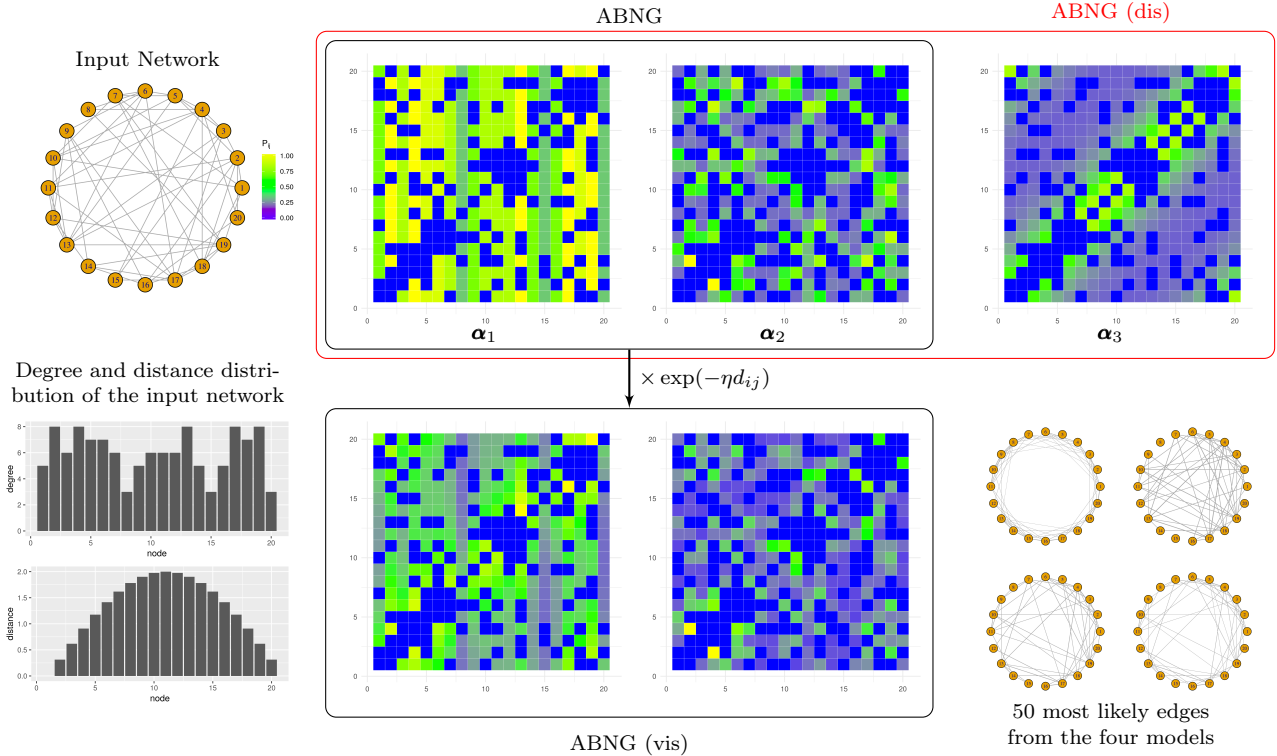
\begin{figure}[!ht]
\centering
\hspace{-.8cm}
\begin{tikzpicture}[node distance = 0.3cm, auto, font=\footnotesize]
    \node [block1] (net) {\includegraphics[width=.22\textwidth]{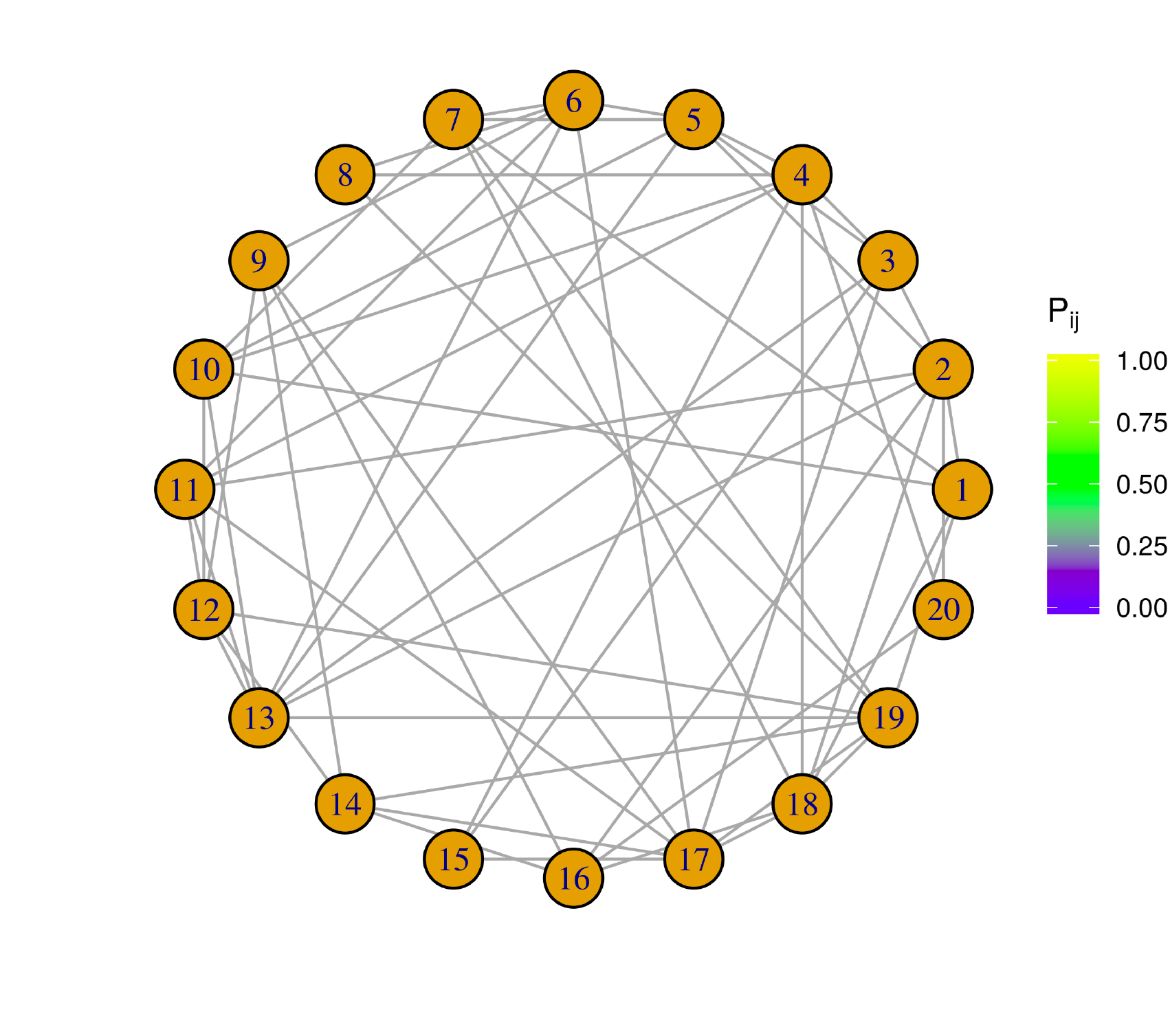}};
    \node [above=-.3cm of net] {Input Network};
    \node [block, right= of net] (abn) {\includegraphics[width=.22\textwidth]{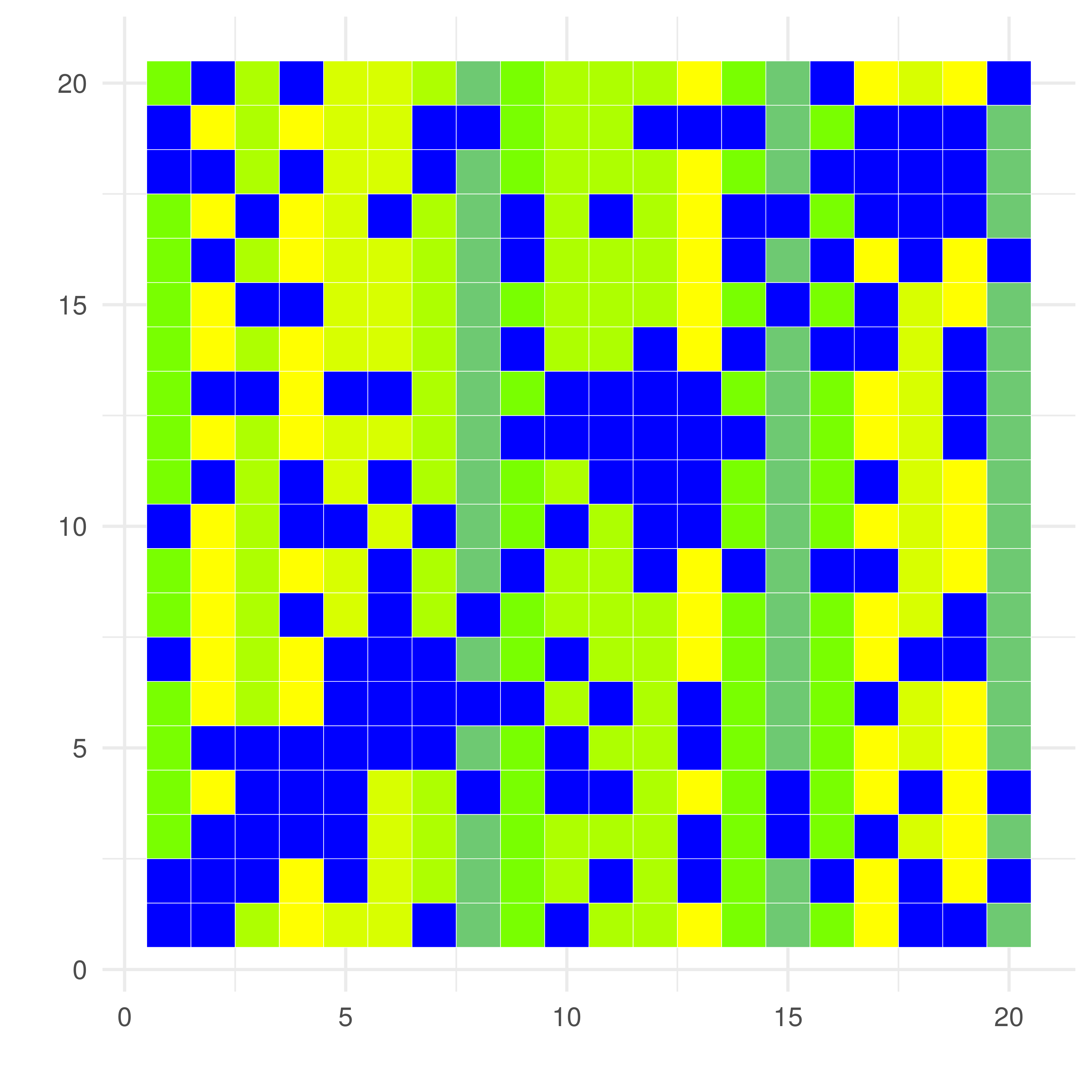}
    \includegraphics[width=.22\textwidth]{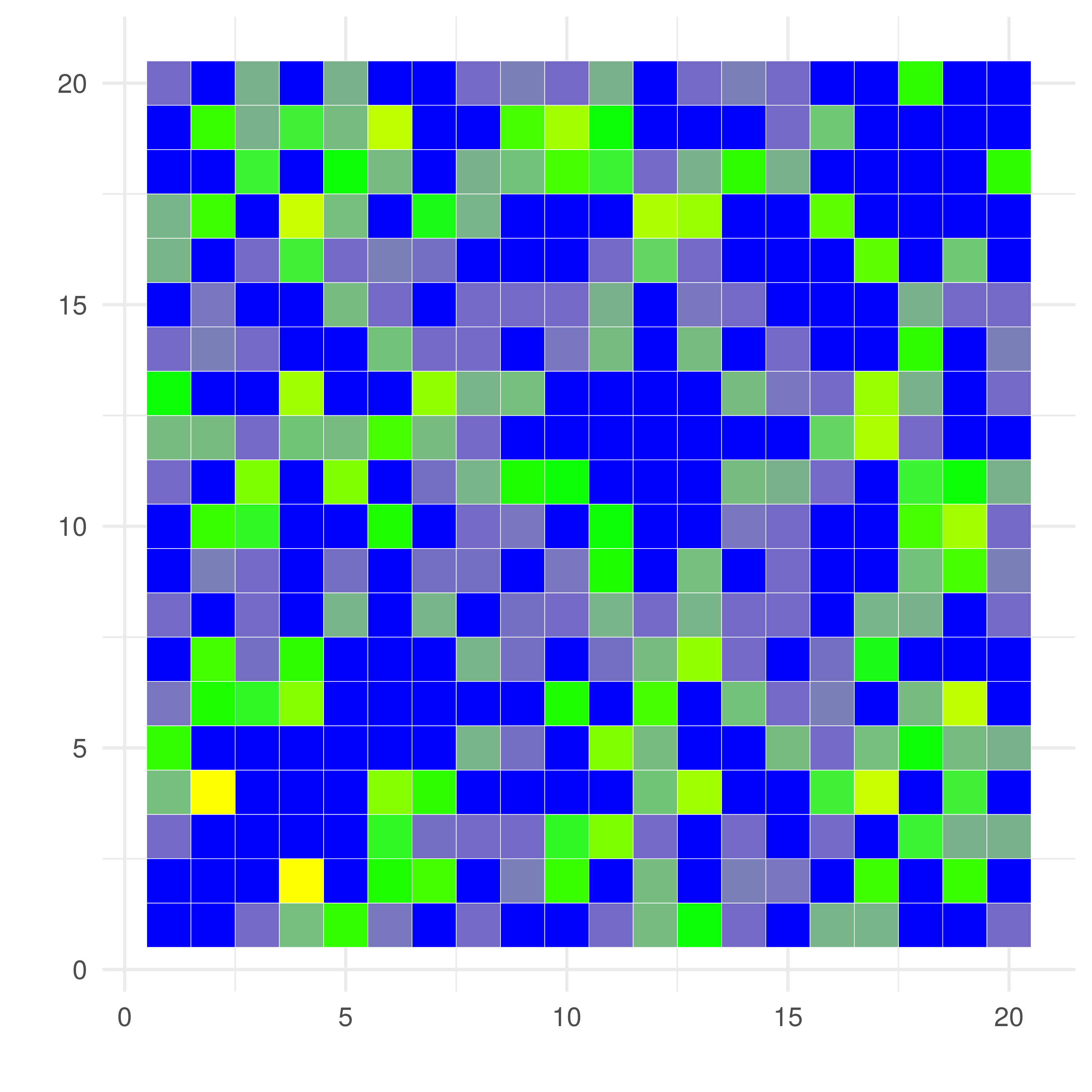}};
    \node [below=-.4cm of abn] {$\pmb{\alpha}_1 \hspace{3.5cm} \pmb{\alpha}_2$};
    \node [above=.1cm of abn] {ABNG};
    \node [block1, right= of abn] (null) {\includegraphics[width=.22\textwidth]{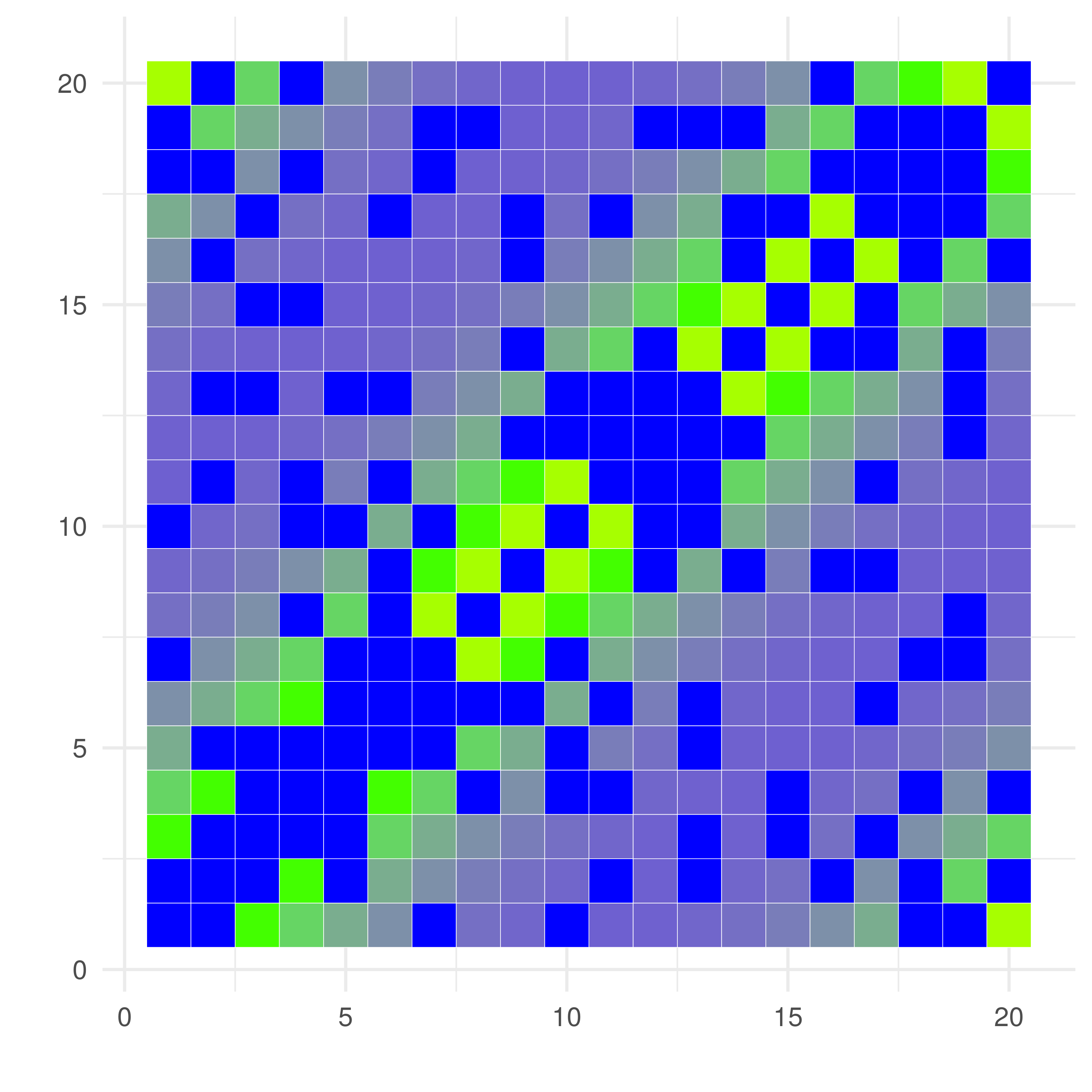}};
    \node [below=-.4cm of null] {$\pmb{\alpha}_3$};
    \node [above=.1cm of null, text=red] {ABNG (dis)};
    \node [draw=red, rounded corners, fit= (null) (abn)] (abnd) {};
    \node [block1, below=1cm of net] (prop) {\includegraphics[width=.22\textwidth]{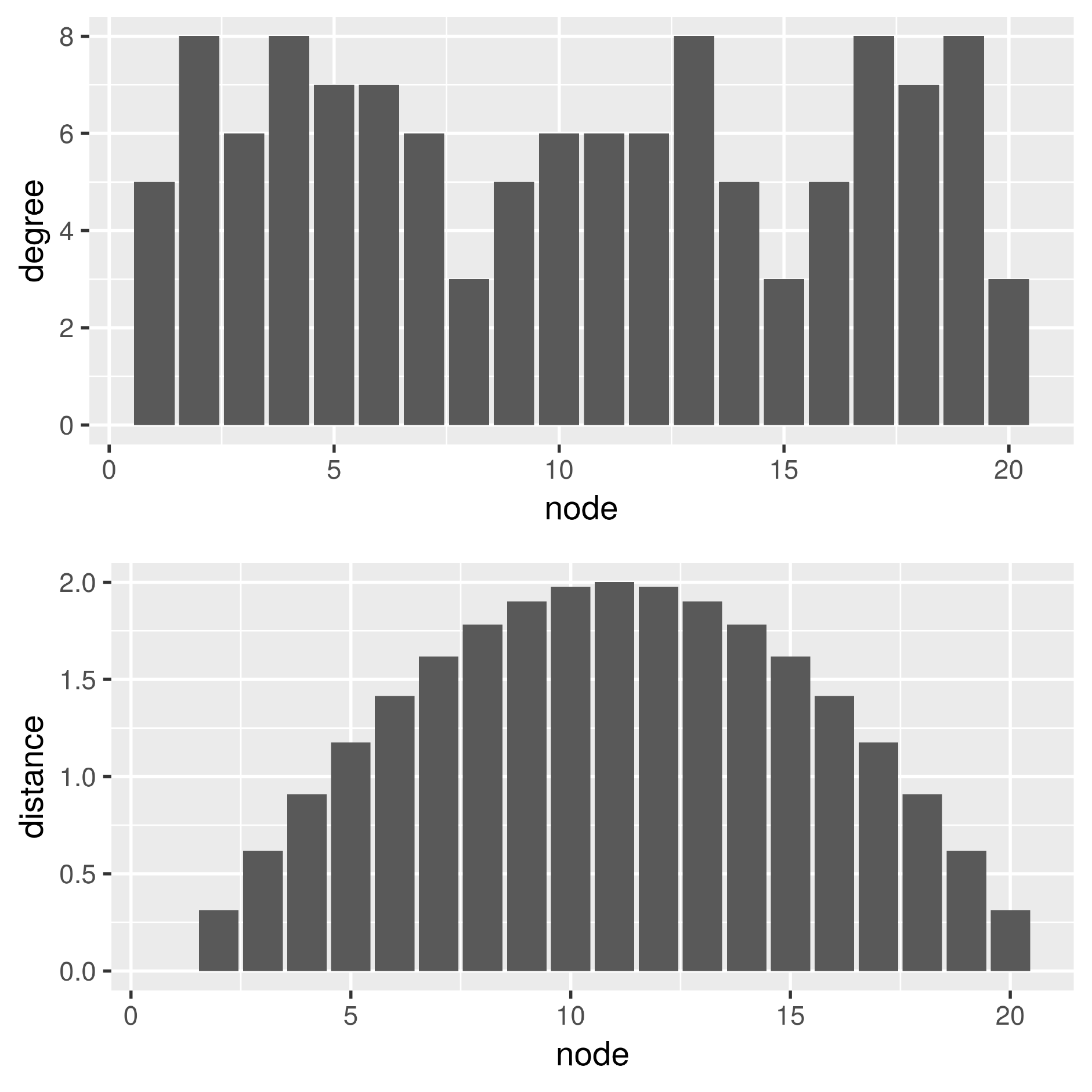}};
    \node [text width=13em, text centered, above=-.1cm of prop] {Degree and distance distribution of the input network};
    \node [block, right= of prop] (abnv) {\includegraphics[width=.22\textwidth]{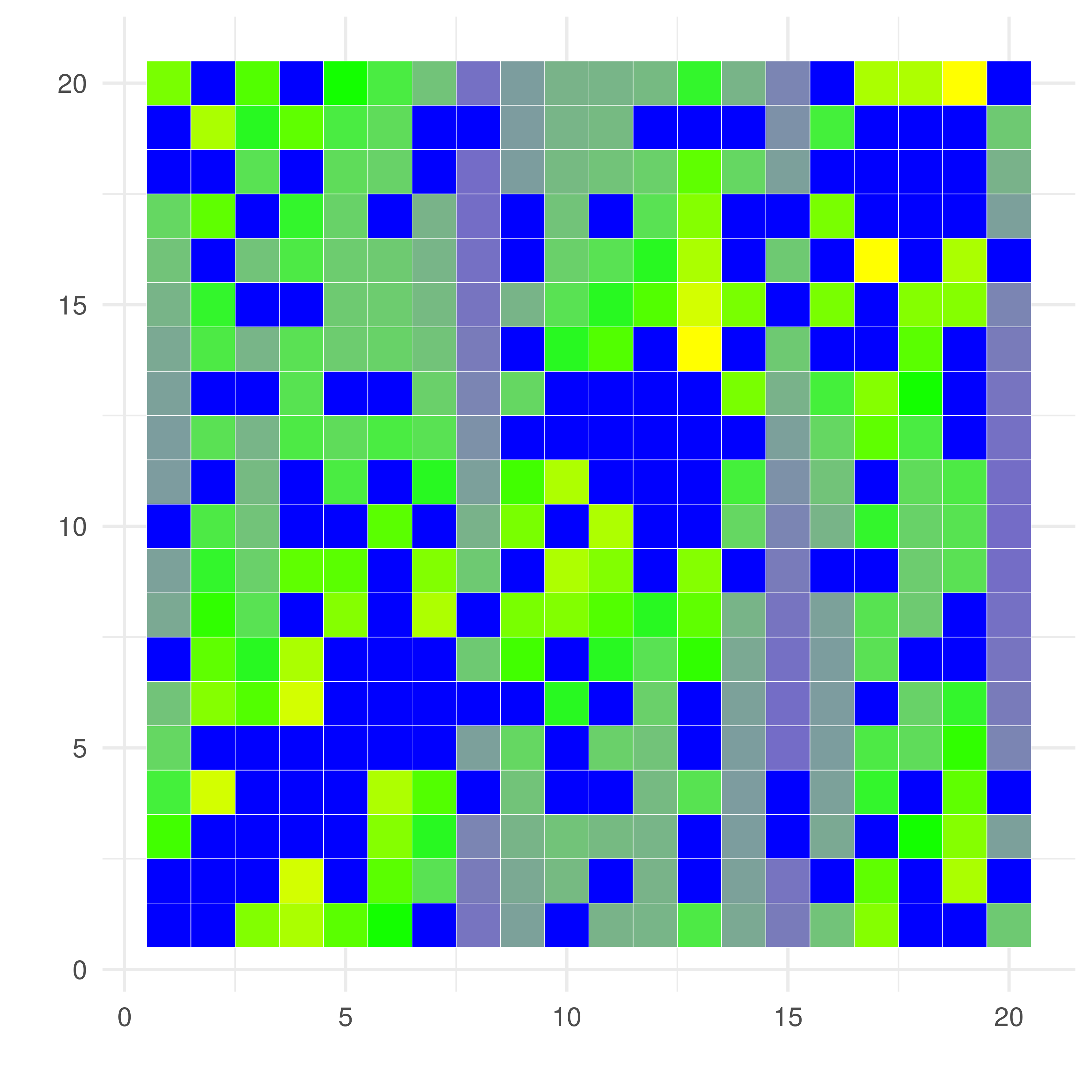}
    \includegraphics[width=.22\textwidth]{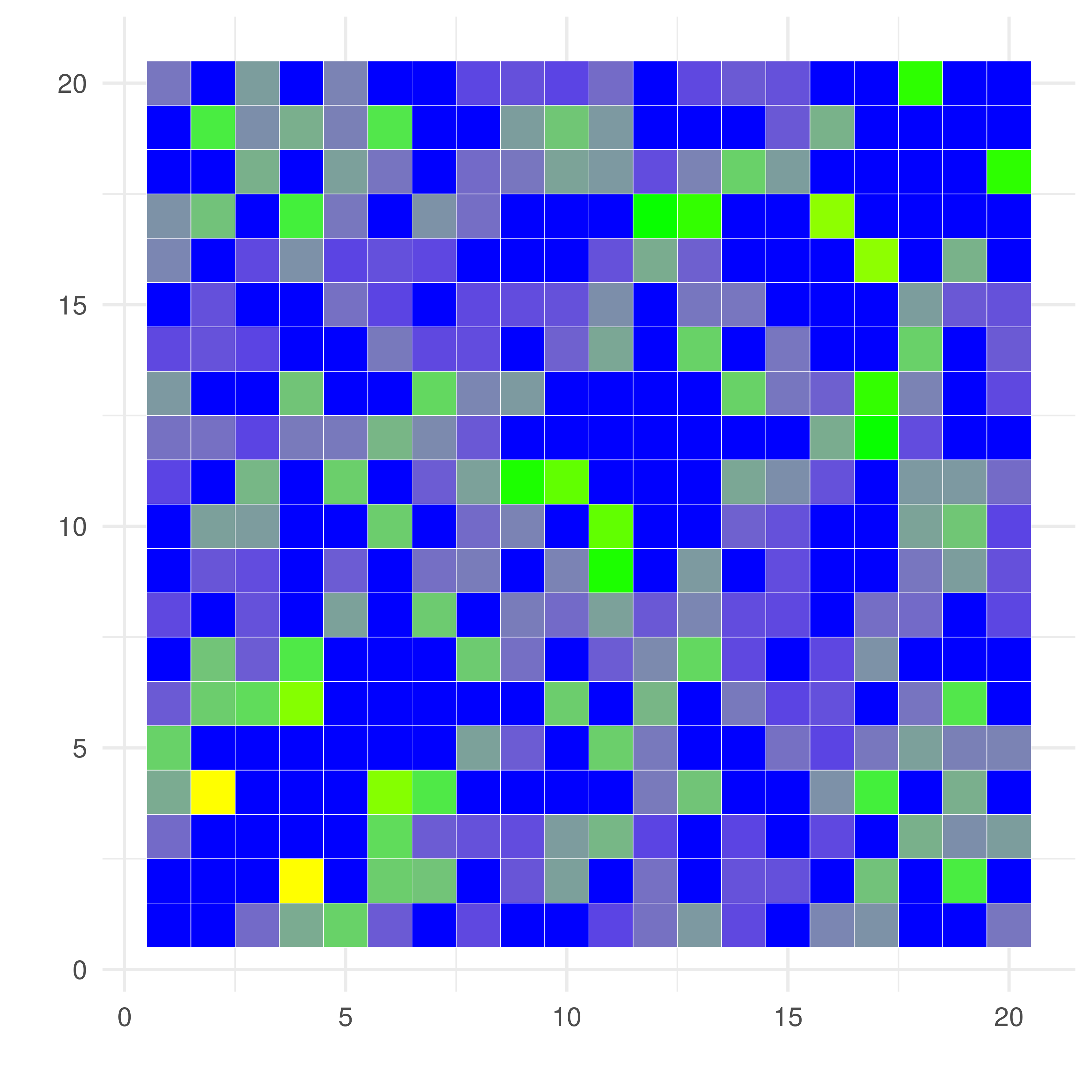}};
    \node [below=.1cm of abnv] {ABNG (vis)};
    \node [block1, right= of abnv] (synth) {\includegraphics[width=.22\textwidth]{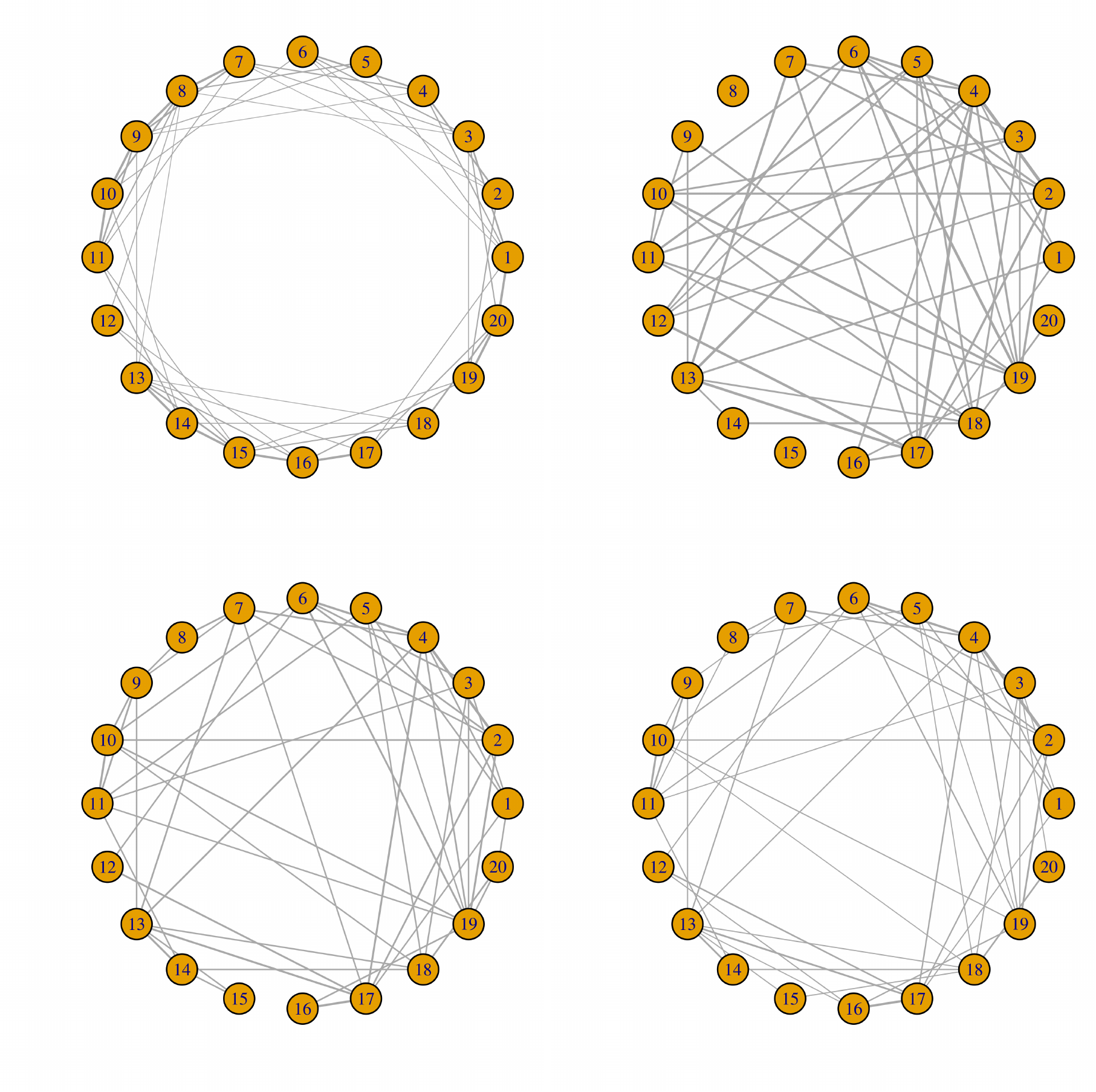}};
    \node [text width=13em, text centered, below=-.3cm of synth] {50 most likely edges from the four models};

    \path [->, thick, line] (abn) -- node[text width=5em] {$\times \exp(- \eta d_{ij})$} (abnv);
\end{tikzpicture}
\caption{Pictorial explanation of how the different generative models work using a toy example: Given the input network, different models use or combine various rules to determine the probability of a new edge. In this toy example, ABNG uses two action: (i) preferential attachment based on degree, and (ii) inverse log-weighted similarity, which leads to output matrices $\pmb{\alpha}_1$ and $\pmb{\alpha}_2$ shown in the black rectangle labelled ABNG. ABNG (dist) also uses an action based on geometric distances shown in matrix $\pmb{\alpha}_3$. The null model uses only $\pmb{\alpha}_3$ to determine probability of a new edge, while ABNG (dist) combines the output of all three actions enclosed in the red rectangle. Finally, ABNG (vis) determines the probability of an edge by multiplying the output of actions in ABNG with a distance penalty term $\exp(- \eta d_{ij})$. To highlight the differences between the four models, we show output networks with 50 most likely edges. The following parameters were used for creating the above example: ABNG $\mathbf{M}=[0.4, 0.6]$, ABNG (dis) $\mathbf{M}=[0.3, 0.3, 0.4]$, ABNG (vis) $\eta=0.5$, null model $\eta=1$.}
\label{fig:models}
\end{figure}

\subsection{Network Construction}
\label{sec:brain_data}
We used DWI data from the 100 unrelated subjects of the HCP 900 subjects data release \cite{VanEssen2013} to get the structural brain networks. The preprocessing of the DWI data to get the corresponding networks is described in detail in \cite{Amico2018}. One network represents the abstracted brain structure of one subject. Nodes in network represent regions of interest (ROIs) in brain and edges represent the density of connecting fibers. All networks share the same set of nodes since brain images of different subjects are regularized into a common template of ROIs. We employed a cortical parcellation into 360 brain regions as recently proposed in \cite{Glasser2016} to produce a structural connectome. Finally, a $\log_{10}$ transformation \cite{fornito2016fundamentals} was applied on the structural connectomes to better account for differences at different magnitudes. Consequently, structural connectivity values ranging between 2.5 and 3.0 were used to produce binarized networks with different edge densities. The results presented in the main text used a cutoff of 3.0 to create binarized networks, while cutoffs of 2.7 and 2.5 were used for dense1 and dense2 networks respectively. To construct the median representative network, the structural connectome was created by taking the median of density of connecting fibres for each pair of ROIs for the 100 subjects, which was then binarized using the previously outlined procedure. Table \ref{tab:stats} shows the statistics and some common network metrics of the brain networks obtained from using different cutoffs for binarization. The modularity is calculated by assigning individual nodes to groups based on their resting state network (RSN) defined using the hierarchical connectivity patterns \cite{ThomasYeo2011}.

\begin{table}[!ht]
\centering
\setlength{\tabcolsep}{4pt}
    \caption{Statistics and network metrics of brain network populations with different edge densities (standard deviation in parentheses).}
    \label{tab:stats}
    \begin{tabular}{lrrrrr}
        \hline\hline
        \multicolumn{1}{c}{Name} & \multicolumn{1}{c}{Average degree} & \multicolumn{1}{c}{Transitivity} & \multicolumn{1}{c}{Assortativity} & \multicolumn{1}{c}{Average path length} & \multicolumn{1}{c}{Modularity} \\
        \hline
        \hline
        main & 6.59 & 0.40 & 0.17 & 5.01 & 0.36 \\
         & (0.22) & (0.01) & (0.04) & (0.25) & (0.01) \\
        \hline
        dense1 & 11.57 & 0.42 & 0.14 & 3.73 & 0.31 \\
         & (0.41) & (0.01) & (0.04) & (0.13) & (0.01) \\
        \hline
        dense2 & 16.02 & 0.43 & 0.11 & 3.22 & 0.28 \\
         & (0.60) & (0.01) & (0.03) & (0.09) & (0.01) \\
         \hline\hline
    \end{tabular}
\end{table}

\subsection{Results: Network modeling}
\label{sec:res-mods}
Another popular class of null models popular in the network science literature are defined by constraining various microscopic properties of a network. The microcanonical ensemble imposes hard constraints on mesoscopic network properties \cite{Cimini2018}. A well-known model that falls in this category is the configuration model \cite{Bender1978, Molloy1995}, where networks are uniformly sampled from an ensemble of networks with a predefined degree distribution. $dk$-random graphs \cite{Orsini2015} further generalize the idea of the configuration model by defining a series of null models or random graph ensembles, where ensemble size is controlled using $dk$-distributions. $dk$-random graphs for $d=0,1,2$ correspond to the random graph model \cite{Erdos1959}, configuration model \cite{Bender1978, Molloy1995} and random graphs with a given joint degree distribution \cite{Stanton2012}, respectively. Randomizing network edges using $dk$-random graphs allows us to check if fixing certain properties in a real-world network can lead to the appearance of other properties as a statistical consequence and if these properties vary significantly from a null model \cite{Betzel2017}.

\begin{figure}[!ht]
    \centering
    \includegraphics[width=\linewidth]{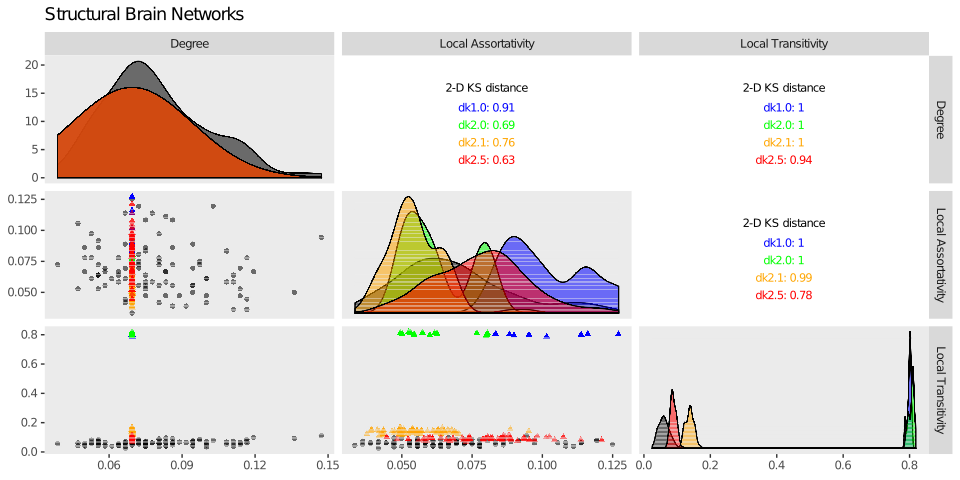}
    \caption{Evaluating the ability of different variants of $dk$-random graphs to reproduce the observed topological variability while using $G^*$ as the input network.}
    \label{fig:sm_dk}
\end{figure}

The $dk$-series methodology allows us to test if the structure of human brain can be simply explained using basic degree or subgraph-based characteristics. Consequently, we use four different variants of $dk$-random graphs as null models to test if the ensembles can capture the topological variability observed in the brain networks. Our results in Figure \ref{fig:sm_dk} show that the $dk$-randomized networks show very little topological variability in the network properties. Although adding average clustering $(dk2.1)$ and degree-dependent clustering $(dk2.5)$ as additional constraints to the random graph ensembles improved the ability of the model to capture the local transitivity, the model is easily outperformed by most versions of the action-based model (see Figure \ref{fig:res}).

\subsubsection{Fine-grain models}
The human brain is known to be spatially organized in a set of specific coherent patterns, called resting state networks (RSNs) \cite{sameshima2014methods}. These RSNs reflect the functional architecture of specific areas, namely: visual, somatomotor, dorsal attention, ventral attention,limbic, frontoparietal, default, and cerebellum \cite{ThomasYeo2011}. Given the prevalence of these hierarchical connectivity patterns, we decided to validate the utility of using this organization in our generative models. This lead to the creation of another variant of the action-based model, ABNG (rsn\_vis), where each RSN of the brain is associated with a different distance penalty parameter. That is, the parameter $\pmb{\eta}$ is now a vector and the distance penalty parameter for a node is determined by its local network organization. Introducing such fine-grained control adds extra degrees of freedom in our models as the distance penalty parameter is optimized separately for each RSN. We also create a similar version of the null model, null (rsn), such that $\pmb{\eta}$ is now a vector of distance penalty terms for each RSN.

\begin{figure}[!ht]
    \centering
    \includegraphics[width=\linewidth]{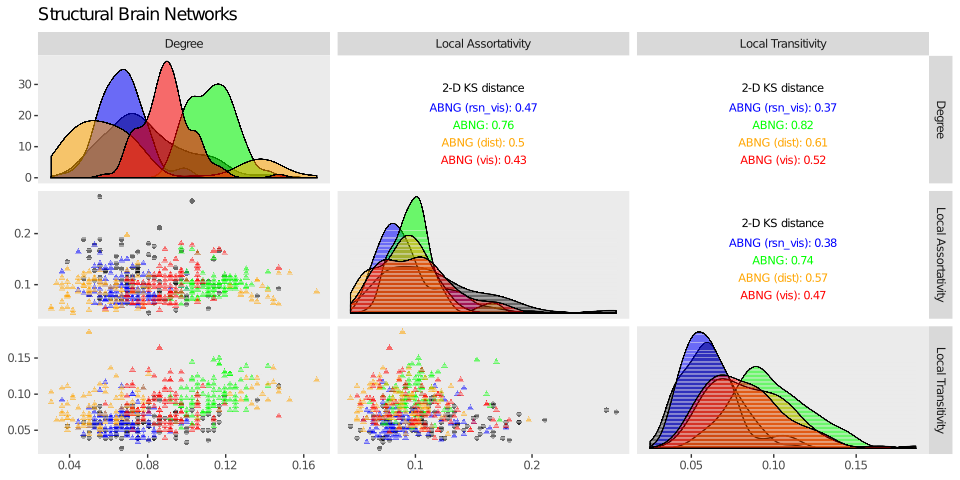}
    \caption{Empirical evaluation of the ability of four variants of ABNG to capture the between-subject variability using the group representative network $G^*$ as the input. ABNG (rsn\_vis) uses separate visibility parameter for each resting state network.}
    \label{fig:sm_rsn}
\end{figure}

As previously, we use the formulation in Equation \ref{eq:genform1} to learn parameters $\pmb{\eta}$ for ABNG (rsn\_vis) using NSGA-II \cite{Deb2002} as the optimization algorithm. Results presented in Figure \ref{fig:sm_rsn} (the null model is replaced by ABNG (rsn\_vis), refer to Section \ref{sec:mod-cv} for instructions on interpreting the results in Figure \ref{fig:res}) show that ABNG (rsn\_vis) marginally outperforms ABNG (vis) (is the improvement enough given the extra degrees of freedom in the model? need to test other network properties). 

\subsubsection{Networks with different densities}
As previously described in Section \ref{sec:brain_data}, binarized brain networks with different edge densities can be obtained by setting different cutoff thresholds during the network construction process. A successful network model should be able to synthesize realistic network irrespective of the threshold used for constructing the networks, which is why we test our models on networks with varying densities.

\begin{figure}[!ht]
    \centering
    \includegraphics[width=\linewidth]{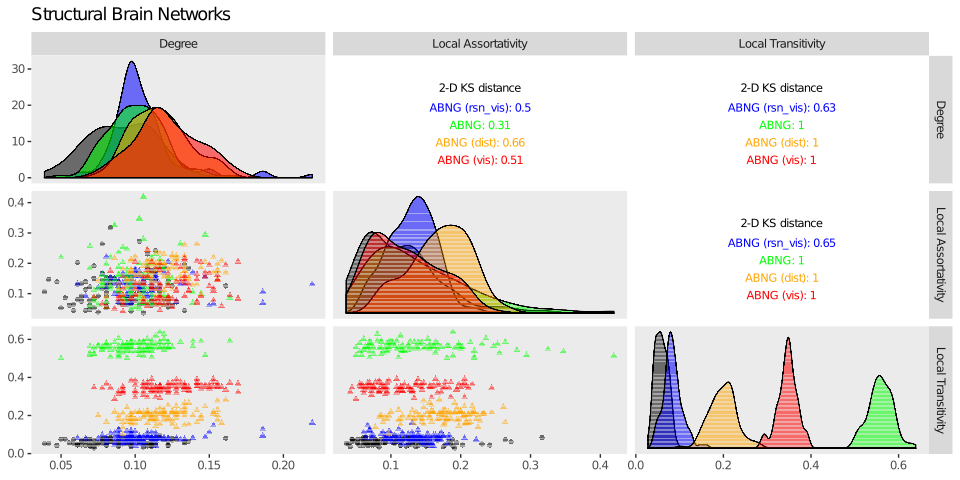}
    \caption{Empirical evaluation of the ability of the aforementioned network models to capture the between-subject variability using the {\it dense1} group representative network the input.}
    \label{fig:sm_rsn-d1}
\end{figure}

\begin{figure}[!ht]
    \centering
    \includegraphics[width=\linewidth]{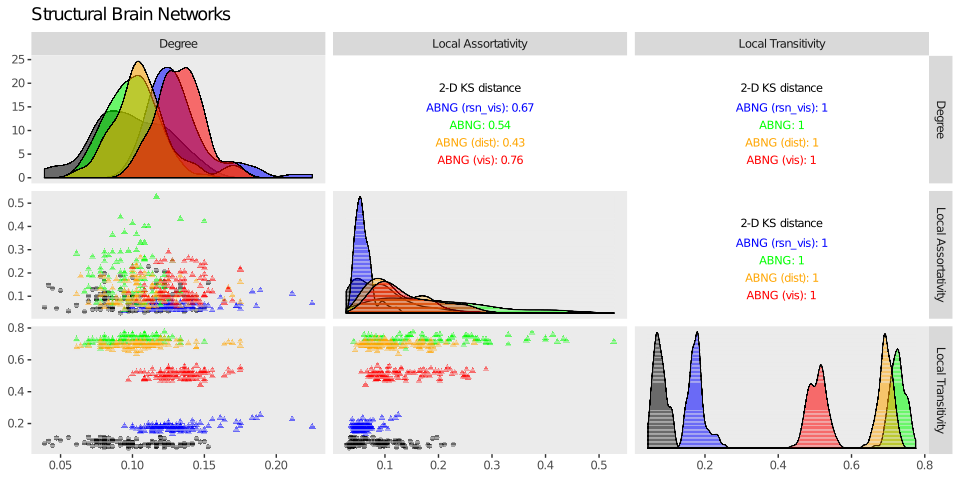}
    \caption{Empirical evaluation of the ability of the aforementioned network models to capture the between-subject variability using the {\it dense2} group representative network as the input.}
    \label{fig:sm_rsn-d2}
\end{figure}

We first construct two more group representative networks (dense1 and dense2) using the procedure described in Section \ref{sec:brain_data}. Structural properties of the network populations created using the new thresholds can be seen in Table \ref{tab:stats}. We learn our models using these networks as input and the results are shown is Figures \ref{fig:sm_rsn-d1} and \ref{fig:sm_rsn-d2}. We observe that the performance of our models degrades, especially in reproducing local transitivity, as the network density increases.




\subsection{Results: Cognitive ability}
\begin{figure}[!ht]
    \centering
    \includegraphics[width=.48\linewidth]{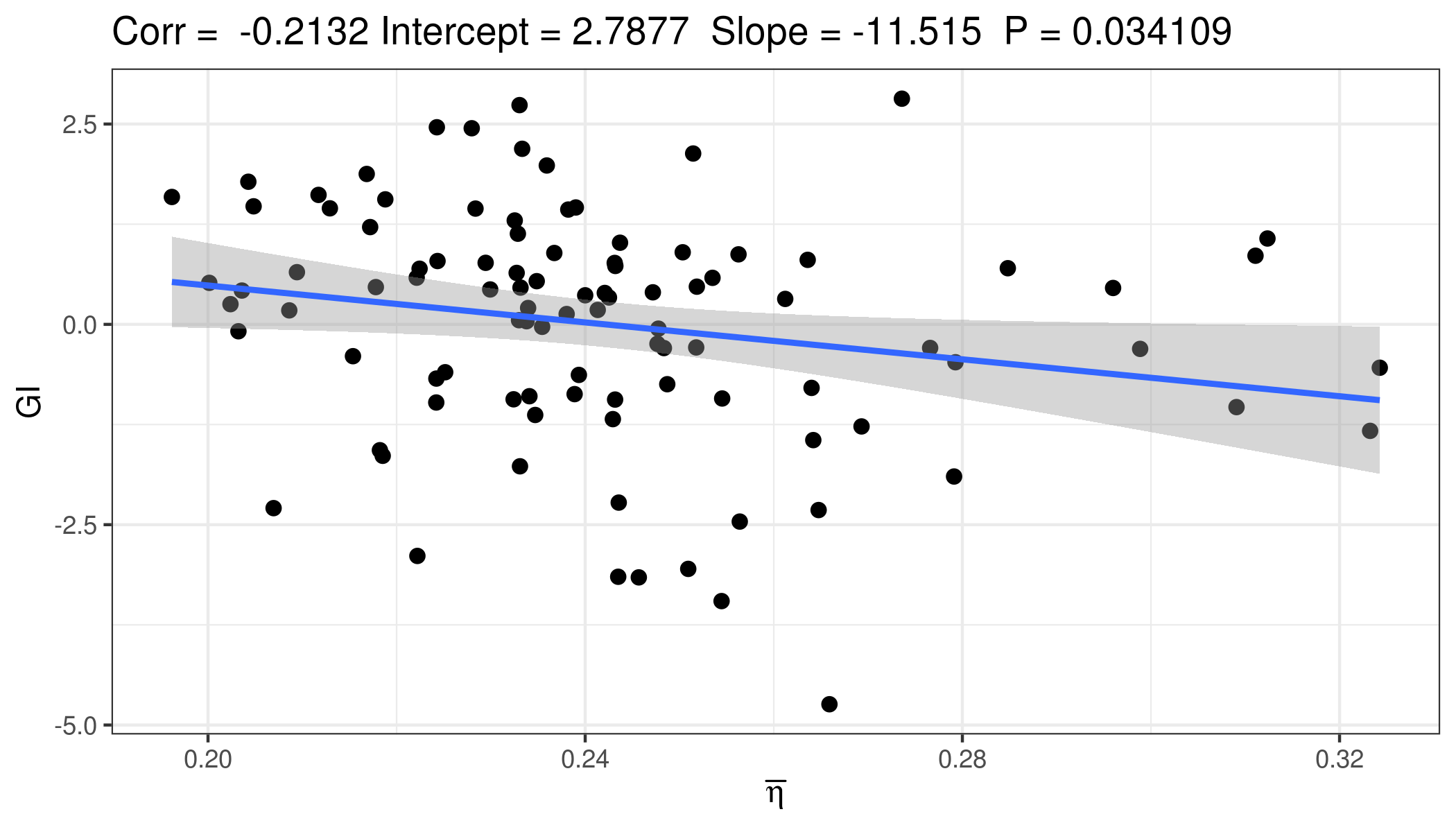}
    \includegraphics[width=.48\linewidth]{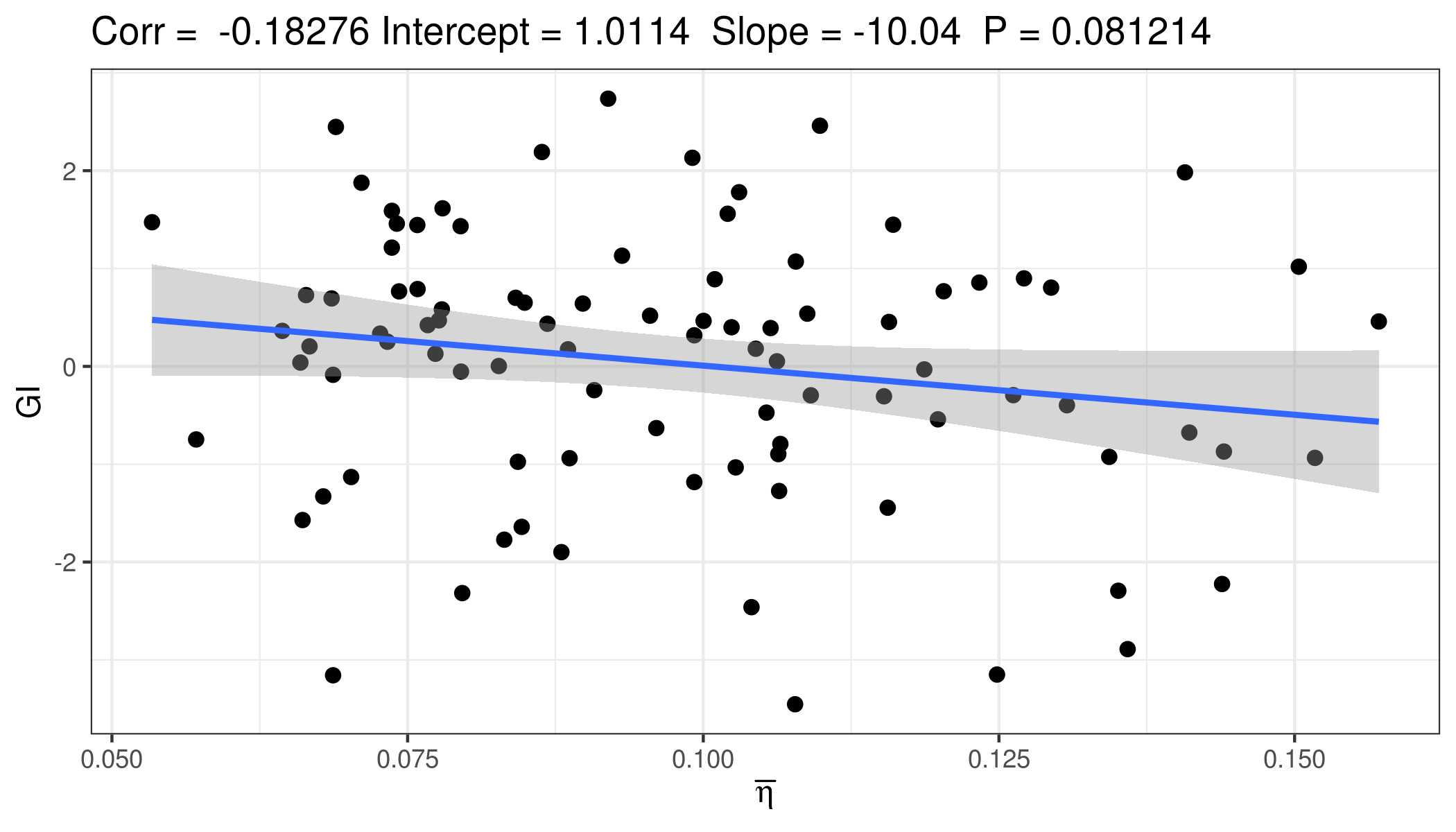}
    \caption{Testing the correlations between mean visibility parameters $\bar{\eta}$ of the individual-based models with the empirically estimated general intelligence of a subject. (left) $\bar{\eta}$ is obtained from the null model, (right) $\bar{\eta}$ is for ABNG (vis), but the same action matrix $\mathbf{M}_{G^*}$ is used for each subject.}
    \label{fig:GI-mods}
\end{figure}

In Section \ref{sec:cog}, we evaluated the ability of the parameters of our best model, ABNG (vis), to provide insights into the cognitive ability of different subjects. Our analysis showed that the mean distance penalty parameter ($\bar{\eta}$) is correlated with the general intelligence of subjects. We can also obtain $\bar{\eta}$ using other individual-based models. We use the null model, and ABNG (vis) where the same action matrix $\mathbf{M}_{G^*}$ was used for each individual but $\eta$'s were separately optimized. Our results in Figure \ref{fig:GI-mods} show that we see the same correlation, albeit a little bit lower, between $\bar{\eta}$ and general intelligence of different individual. These results further show that ABNG (vis) in addition to capturing the structure of the networks provides the best insights into other aspects of the human brain.

\begin{figure}[!ht]
    \centering
    \includegraphics[width=.48\linewidth]{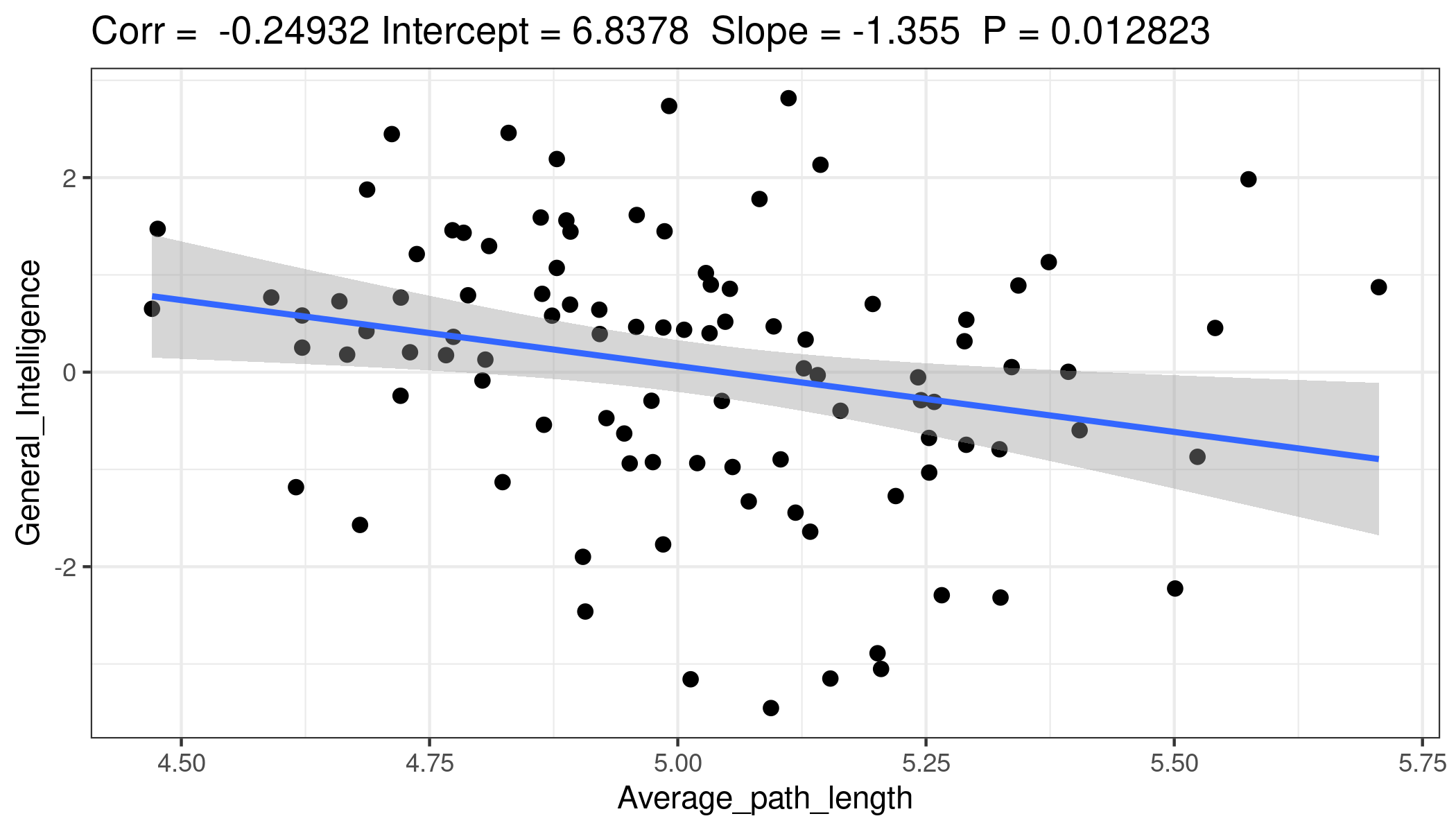}
    \includegraphics[width=.48\linewidth]{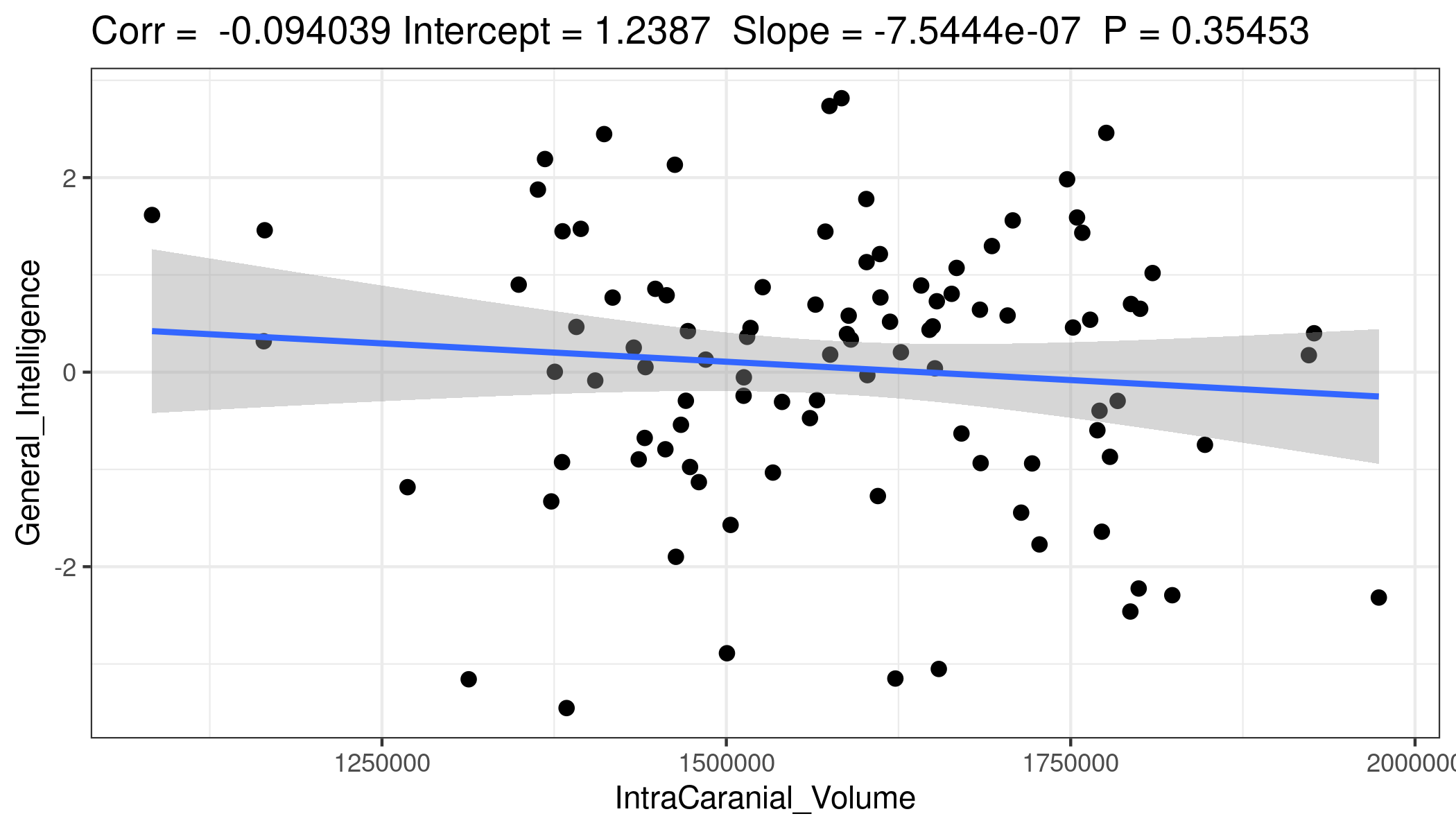}
    \caption{Testing the correlations between properties of the brain of an individual, average path length (left) and intra-cranial volume, with the empirically estimated general intelligence of a subject. }
    \label{fig:GI-br-prop}
\end{figure}

We also evaluated the correlations between $\bar{\eta}$ for the individual-based model ABNG (vis) and for six different measures of cognitive ability described below. In Figure \ref{fig:cog}, the correlations and p-values are also reported. Using a significance level of $\alpha=0.05$, we can conclude that the correlation is not insignificant for four of the cognitive ability measures. In our evaluation of the cognitive ability, we used the following six measures:

\begin{figure}[!ht]
    \centering
    \includegraphics[width=.48\linewidth]{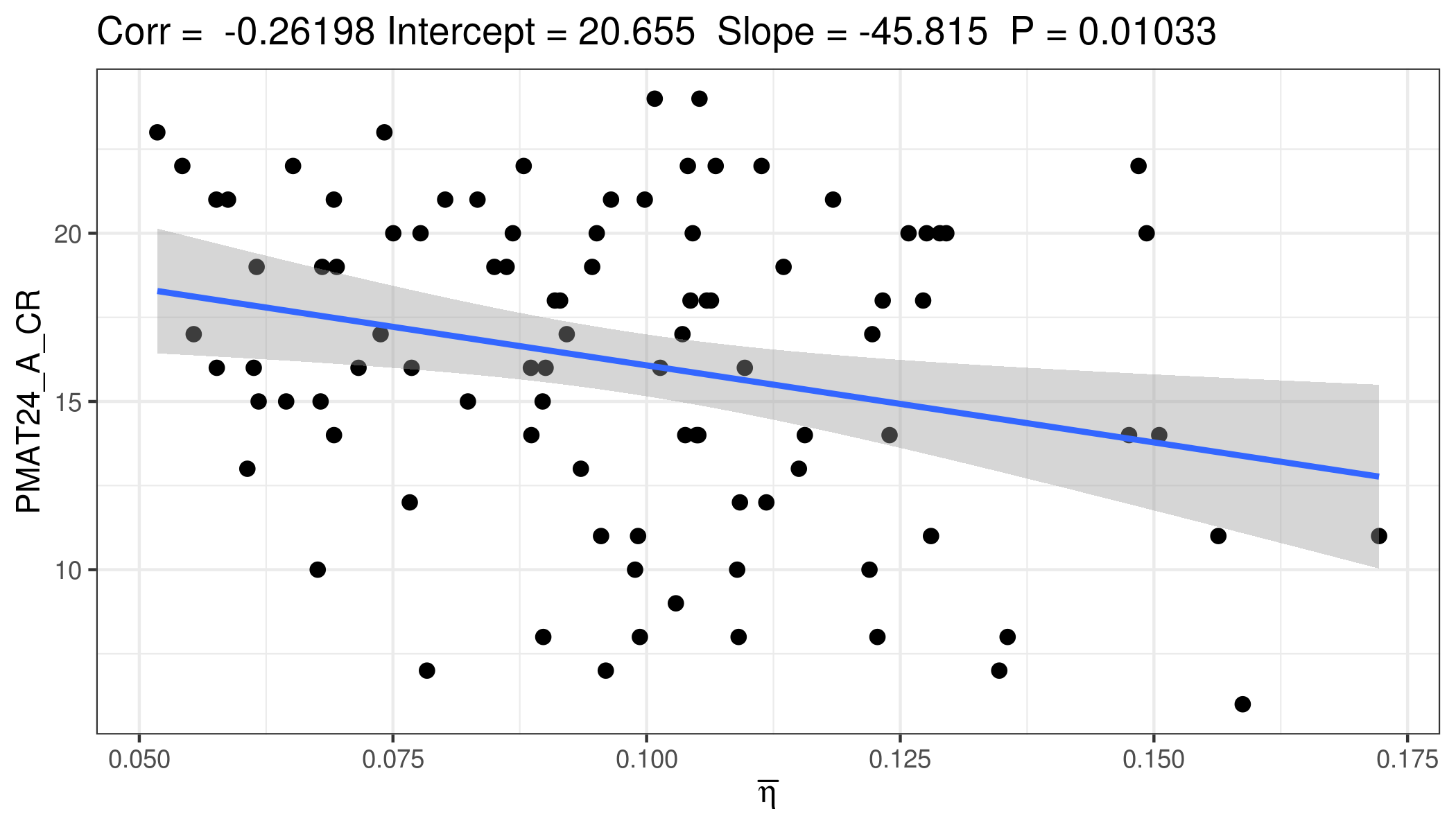}
    \includegraphics[width=.48\linewidth]{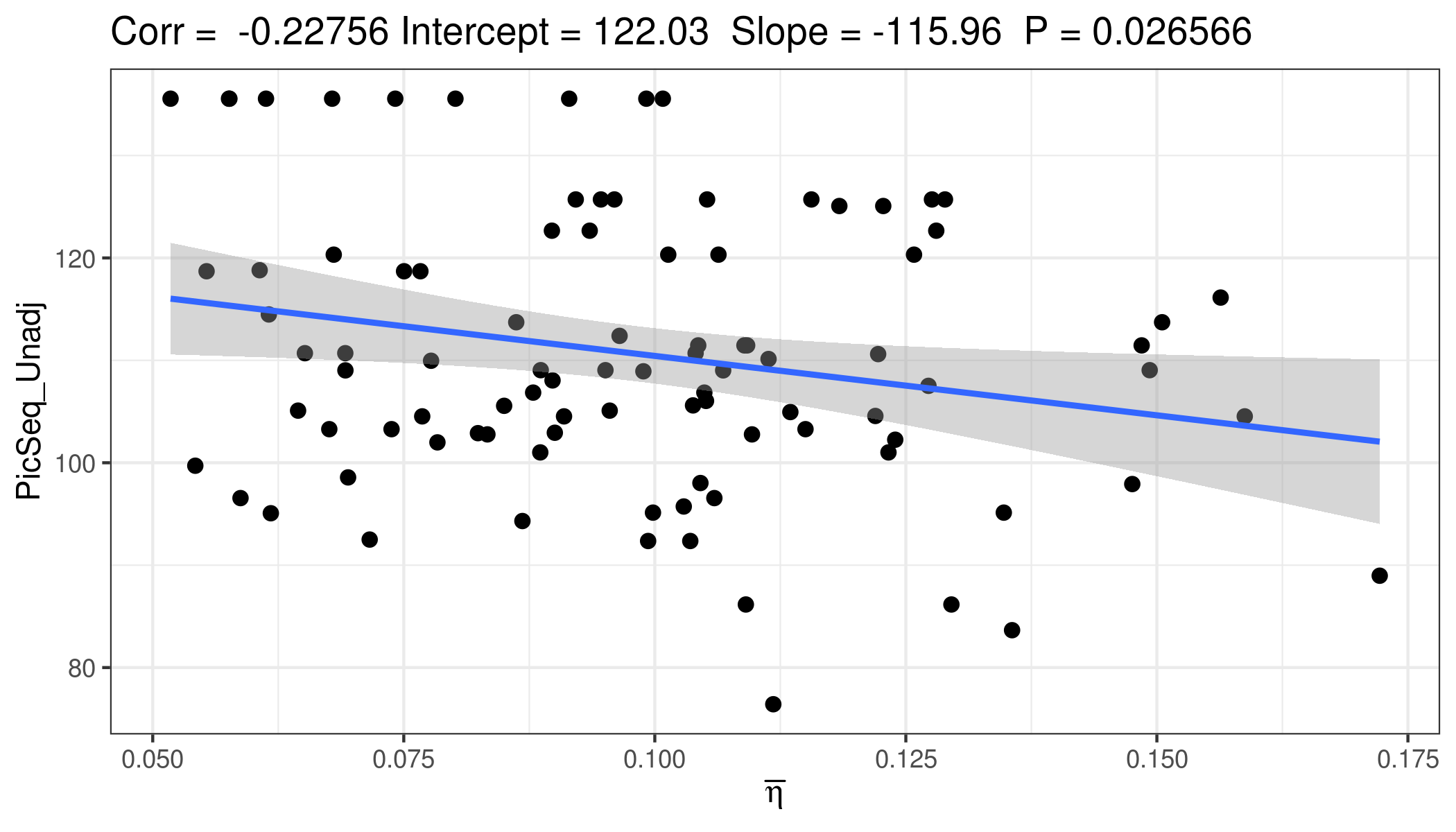}\\
    \includegraphics[width=.48\linewidth]{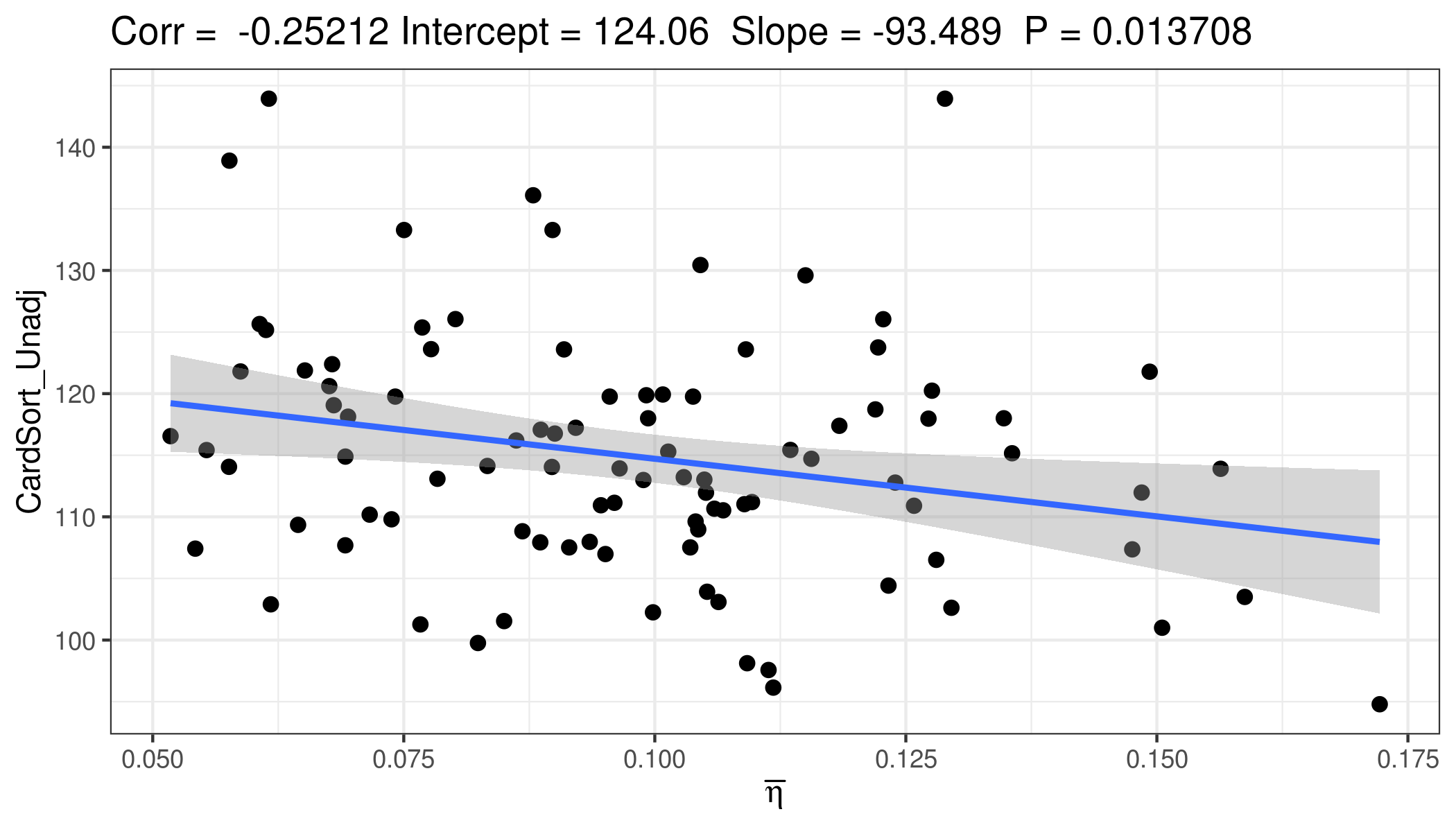}
    \includegraphics[width=.48\linewidth]{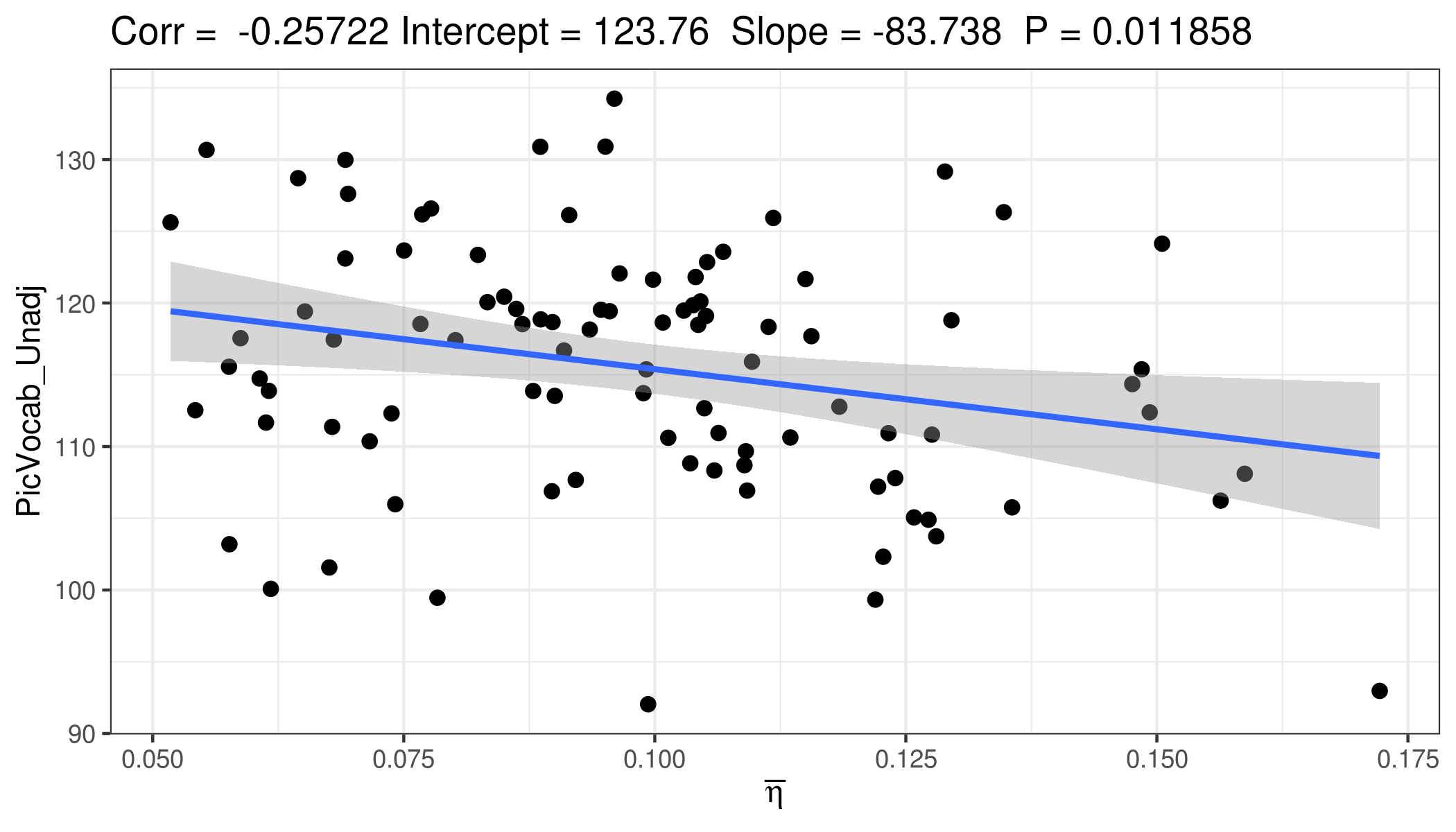}\\
    \includegraphics[width=.48\linewidth]{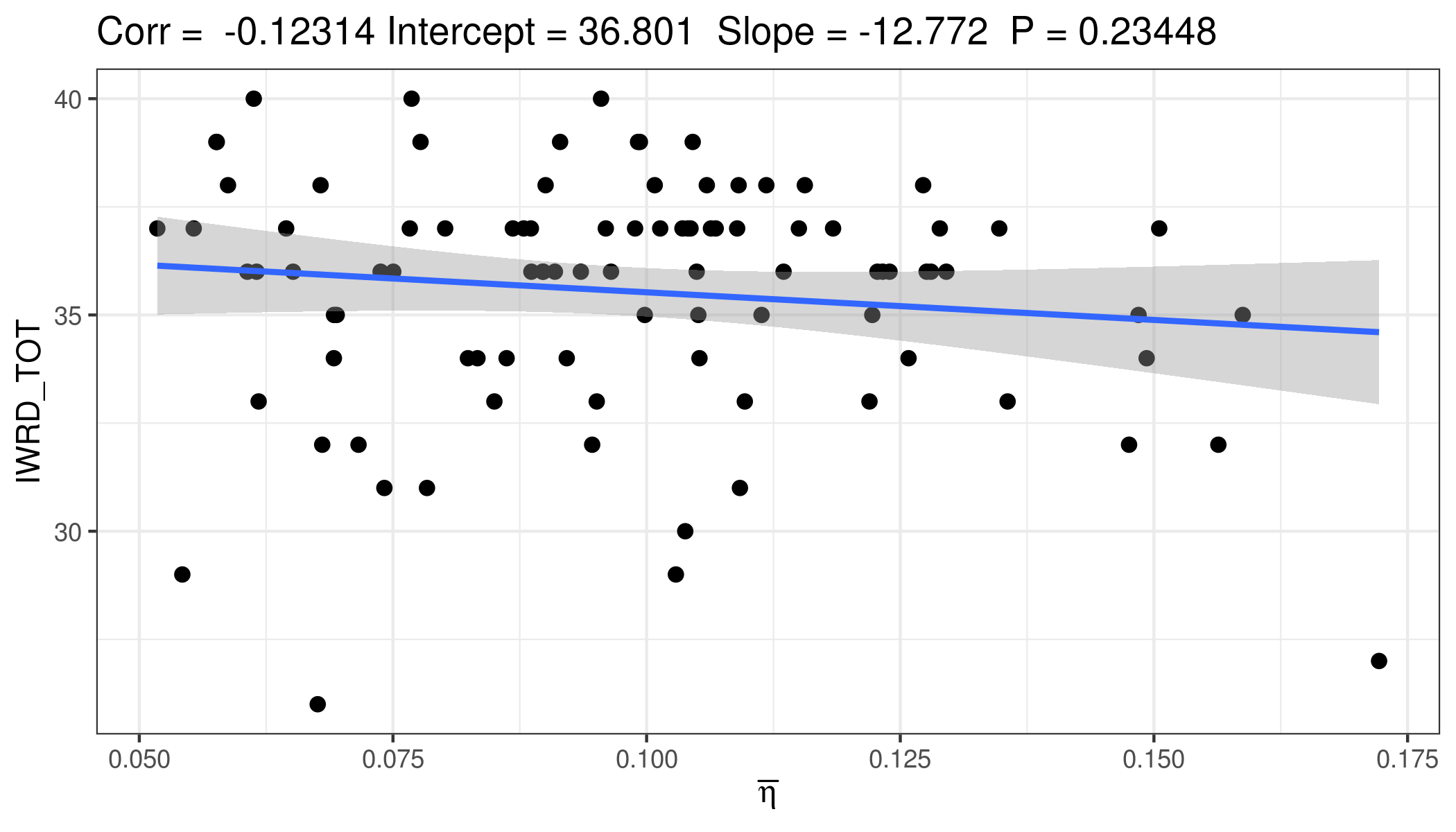}
    \includegraphics[width=.48\linewidth]{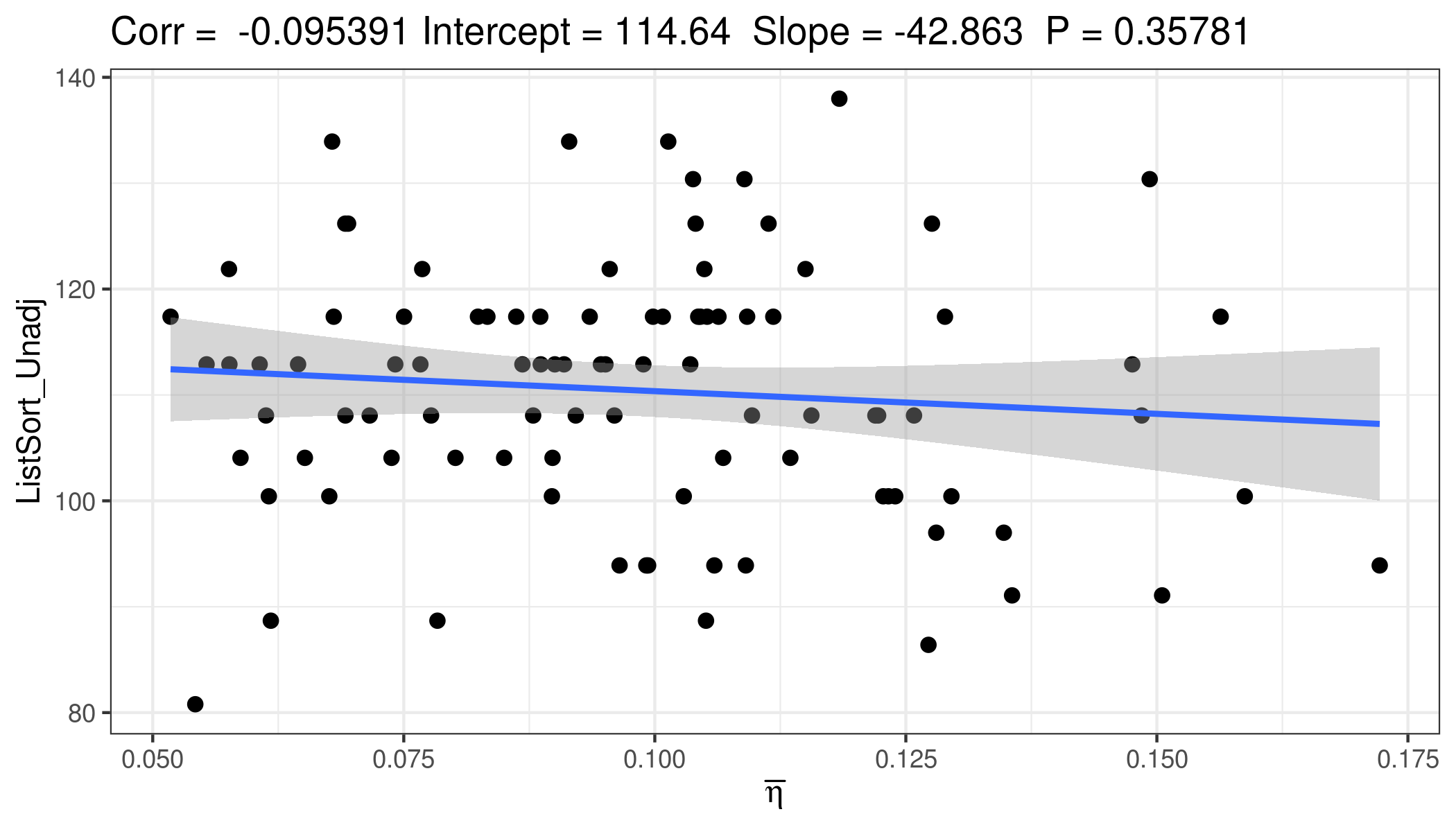}
    \caption[Correlation of $\bar{\eta}$ with various measures of cognitive ability]{Correlation of $\bar{\eta}$ with various measures of cognitive ability}
    \label{fig:cog}
\end{figure}

\begin{enumerate}
    \item {\bf Fluid Intelligence} (PMAT24\_A\_CR): It reflects general cognitive ability, especially as it relates to cognitive control. It is also related to novel/flexible processing.
    \item {\bf Episodic Memory} (PicSeq\_Unadj): It involves recalling increasingly lengthy series of illustrated objects and activities that are presented in a particular order on the computer screen. The participants are asked to recall the sequence of pictures that is demonstrated over two learning trials; sequence length varies from 6-18 pictures, depending on age. 
    \item {\bf Cognitive Flexibility} (CardSort\_Unadj): Two target pictures are presented that vary along two dimensions (e.g., shape and color). Participants are asked to match a series of bivalent test pictures (e.g., yellow balls and blue trucks) to the target pictures, first according to one dimension (e.g., color) and then, after a number of trials, according to the other.
    \item {\bf Language and Vocabulary Comprehension} (PicVocab\_Unadj): This measure of receptive vocabulary is administered in a computerized adaptive format. The respondent is presented with an audio recording of a word and four photographic images on the computer screen and is asked to select the picture that most closely matches the meaning of the word.
    \item {\bf Verbal Episodic Memory} (IWRD\_TOT): Participants are shown 20 words and asked to remember them for a subsequent memory test. They are then shown 40 words (the 20 previously presented words and 20 new words matched on memory related characteristics). They decide whether they have seen the word previously by choosing among ``definitely yes'', ``probably yes'', ``probably no'', and ``definitely no''.
    \item {\bf Working Memory} (ListSort\_Unadj): This task assesses working memory and requires the participant to sequence different visually- and orally-presented stimuli.
\end{enumerate}

\begin{figure}[!ht]
    \centering
    \includegraphics[width=.75\linewidth]{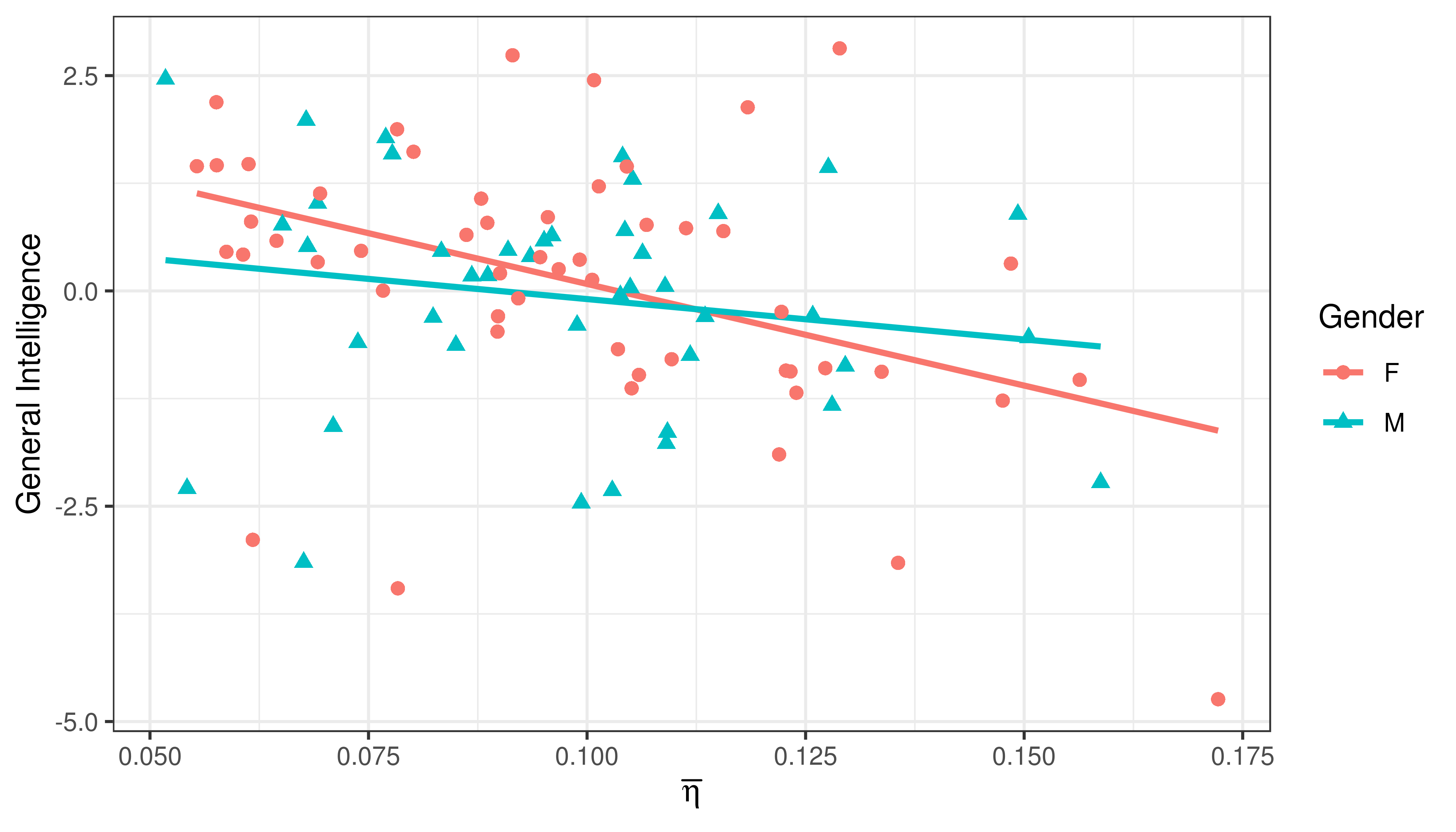}
    \caption[Correlation between model parameters and general intelligence]{The plot shows results for correlation between general intelligence and $\bar{\eta}$ ($r=-0.318$, $r_M=-0.178$, $p_M=0.259$ and $r_F=-0.429$, $p_F=0.001$), where F and M are gender of the subjects.} 
    \label{fig:GI-gen}
\end{figure}

Given the important role of gender in neuroscience \cite{Cahill2006}, we modify Figure \ref{fig:GI} to explicitly account for the gender of a subject as a factor in Figure \ref{fig:GI-gen}. We observe that there are significant differences in the correlations between the two groups, thus warranting further investigation.

\bibliographystyle{unsrt}
\bibliography{library}

\end{document}